\documentclass[sigconf,nonacm]{acmart} %
\usepackage[frozencache,cachedir=.]{minted}
\usepackage[framemethod=TikZ]{mdframed}

\usepackage{fontawesome5}

\usepackage{colortbl}

\usepackage{subfig}
\usepackage{float}
\usepackage{graphics}
\usepackage{arydshln}
\usepackage{hhline}
\usepackage{tikz}
\usepackage{pifont}%
\usepackage{minted}
\usepackage{booktabs}
\usepackage{makecell}
\usepackage{enumitem}
\usepackage{multirow}

\AtBeginDocument{%
  }

\definecolor{tableBlue}{RGB}{109, 155, 195}

\definecolor{backcolour}{rgb}{0.95,0.95,0.92}
\definecolor{darkblue}{RGB}{0,0,139} %
\definecolor{darkgreen}{RGB}{0,100,0} %
\definecolor{brownishorange}{RGB}{204,85,0} %

\definecolor{new-green}{rgb}{0.104,0.667,0.229}

\newcommand{\promptbox}[2]{%
  \noindent\colorbox{black!5}{%
    \parbox{\dimexpr\linewidth-2\fboxsep\relax}{%
      \textbf{#1}\par\small\ttfamily #2
    }%
  }\par\vspace{0.6em}%
}

\setcopyright{acmlicensed}
\copyrightyear{2025}
\acmYear{2025}
\acmDOI{XXXXXXX.XXXXXXX}

\acmConference[CHI '26]{Make sure to enter the correct
  conference title from your rights confirmation email}{13--17 April,
  2026}{Barcelona, Spain}
\acmBooktitle{Proceedings of the ACM (Association of Computing Machinery) CHI conference on Human Factors in Computing Systems (CHI '26), 13--17 April, 2026, Barcelona, Spain}
\acmISBN{978-1-4503-XXXX-X/18/06}

\begin{document}
\title[Reading.help]{Reading.help: Supporting EFL Readers with Proactive and On-Demand Explanation of English Grammar and Semantics}

\author{Sunghyo Chung}
\affiliation{%
  \institution{Kakao}
  \city{Seongnam}
  \country{South Korea}}
\email{shawn.hyo@kakaocorp.com}
\orcid{0000-0002-4934-9073}

\author{Hyeon Jeon}
\affiliation{%
  \institution{Seoul National University}
  \city{Seoul}
  \country{South Korea}}
\email{hj@hcil.snu.ac.kr}
\orcid{0000-0002-9659-2922}

\author{Sungbok Shin}
\authornote{Corresponding Author.}
\affiliation{%
  \institution{Aarhus University}
  \city{Aarhus}
  \country{Denmark}}
\email{sbshin90@cs.umd.edu}
\orcid{0000-0001-6777-8843}

\author{Md Naimul Hoque}
\affiliation{%
  \institution{University of Iowa}
  \city{Iowa City}
  \country{United States}}
\email{nhoque@umd.edu}
\orcid{0000-0003-0878-501X}

\renewcommand{\shortauthors}{Chung et al.}
\newcommand{\techname}{\textsc{Reading.help}}
\begin{abstract}

A large portion of texts is written in English, but readers who see English as a Foreign Language (EFL) often struggle to read texts accurately and swiftly. EFL readers seek help from professional teachers and mentors, which is limited and costly. In this paper, we explore how an intelligent reading tool can assist EFL readers. We conducted a case study with EFL readers in South Korea. We at first developed an LLM-based reading tool based on prior literature. We then revised the tool based on the feedback from a study with 15 South Korean EFL readers. The final tool, named \techname{}, helps EFL readers comprehend complex sentences and paragraphs with on-demand and proactive explanations. We finally evaluated the tool with 5 EFL readers and 2 EFL education professionals. Our findings suggest \techname{} could potentially help EFL readers self-learn English when they do not have access to external support.

\end{abstract}



\begin{teaserfigure}
    \includegraphics[width=\textwidth]{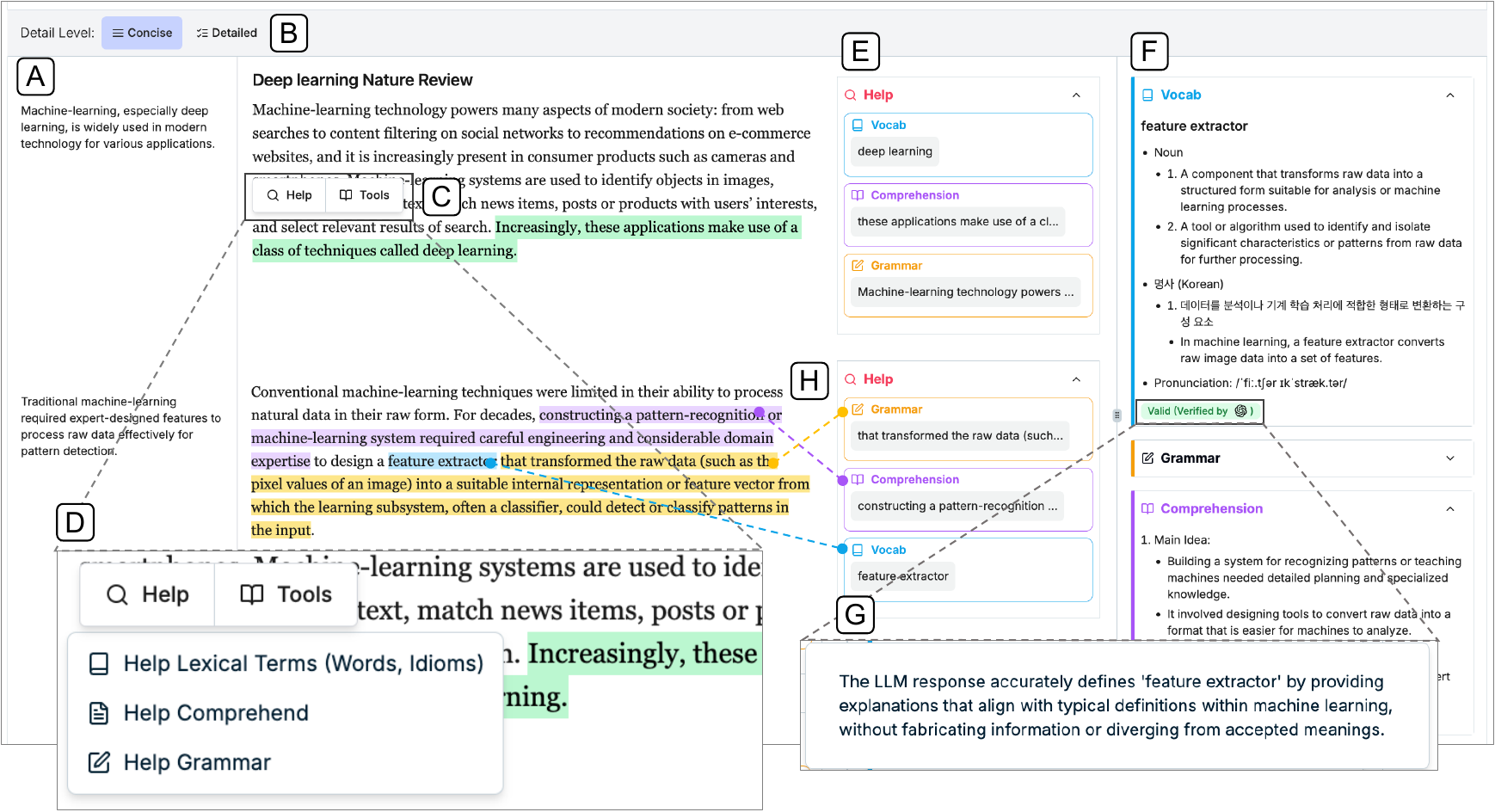}
    \caption{\textbf{The \techname{} interface.} \techname{} assists English as a Foreign Language (EFL) readers in understanding English texts by identifying areas that may confuse readers. The system assists EFL readers with (A) content summaries and (B) adjustable level of summaries (concise or detailed). (C) When users select a specific text, they can access the supporting tools. (D) The Tools menu expands to offer three assistance options: Lexical Terms, Comprehension, and Grammar. (E) The system identifies potentially challenging content for EFL readers and provides recommendations for each paragraph. (F) When the vocabulary tool is selected, the explanations appear with definitions and contextual information. (G) All explanations go through a validation process through a second LLM, which validates the reasoning of the first LLM. (H) The suggestions are linked to the text through highlighting.}
    \Description{Figure 1 displays the Reading.help interface designed for EFL readers. The interface features several components: (A) a section providing content summaries; (B) an option to adjust the summary’s level (concise or disabled); (C) a panel for selecting specific texts to access supporting tools; (D) an expandable Tools menu offering Lexical Terms, Comprehension, and Grammar options; (E) highlighted sections where the system identifies potentially challenging content and provides paragraph-level recommendations; (F) a vocabulary tool that shows definitions and contextual explanations; (G) a validation process where a second LLM checks the initial reasoning; and (H) highlighted text linking the suggestions to the original content.}
    \label{fig:teaser}
\end{teaserfigure}

\maketitle

\section{Introduction}
\label{sec:01_intro}

English is the most widely used language for communication in academic, professional, and social contexts worldwide. A vast amount of literature, including books and documents, is written in English, with millions of new documents being created at any given moment. As a result, the ability to read English accurately and efficiently remains a crucial skill for individuals globally. Acknowledging this need, many educational institutions in countries where English is considered a foreign language incorporate English-language curricula at various levels, such as in international schools with O and A levels~\cite{o_levels}. However, these educational programs are often limited and costly, remaining inaccessible to a significant portion of the population. Even students who receive formal English education face difficulties due to the lack of exposure to advanced vocabulary and cultural references outside the classroom.

Due to the challenges in formal English education, many readers who consider English a foreign language (EFL) seek resources to self-educate themselves. 
There are now online courses, but they are limited in scope and do not provide on-demand and situational support readers need while reading an article. 
Various tools are available to fill this gap, including electronic dictionaries and online translators such as the Google Translate\footnote{\url{https://translate.google.com/}}.  
While electronic dictionaries are widely used, they offer limited support in identifying areas where readers may misinterpret or misunderstand the text.
Online translators offer a convenient means of understanding English texts. 
While their strategic use (e.g., searching for short phrases, clauses, etc.) can support both comprehension and learning, providing full-text translations may lead many non-native English learners to translate entire passages, increasing reliance on the tool and reducing motivation to engage with new vocabulary or syntactic structures~\cite{afiliani2024use}.
These points suggest that EFL readers could benefit from a tool that supports independent interpretation of English texts as well as offer learning opportunities for the readers.

With this objective in mind, we introduce \techname{}, an interface designed to help EFL readers' independent reading of English texts by leveraging natural language processing (NLP) techniques.
\techname{} mainly targets EFL readers, who have university degrees, and are willing to independently interpret English texts. 
\techname{} contains two different modules: 1) an \textbf{identification} module for detecting words and sentences that an EFL reader may find difficult; and 2) an \textbf{explanation} module for clarifying vocabulary, grammar, and overall context of the text. We developed two specialized and interpretable neural models for detection challenging words and sentences. We then use a Large Language Model (LLM) to provide explanations for the identified words and phrases. Users of the tool can also manually highlight any text and request explanation. To ensure the reliability of these explanations, a secondary Large Language Model (LLM) evaluates and refines the information provided. Finally, the explanation module generates concise summaries of paragraphs within the text, helping readers to grasp the main idea without becoming overwhelmed by details.

As a case study, we base our work in the context of South Korea. 
We initially conducted a pilot study with 15 South Korean EFL readers. 
Based on the feedback from the pilot study, we revised the tool and subsequently evaluated \techname{} in a study involving 5 EFL readers and 2 domain experts in English education for EFL readers in South Korea. 
We assess the effectiveness of our tool and observe how these readers benefit from \techname{}. 
We find that people use our tool mainly to address text complexity issues and gain comprehension support. 
\techname{} reports an average recall of 87\% when recommending three elements per dimension.  
Furthermore, while the self-validation performance of LLMs is comparable to that of humans in detecting incorrect meaning issues, LLMs remain limited in identifying other issues, such as missing elements.

\section{Background and Related Work}
\label{sec:02_background}

We explain the background of our work in three parts: (1) difficulties for EFL readers, (2) computational approaches to assist academic reading, and (3) work that augments writing. 

\subsection{Difficulties in EFL Reading} %
\label{subsec:ESL-difficulties}

We identified studies to understand effective methods to learn English, in both reading~\cite{carr1981readingability, graesser1920readingcomponents, iwai2008perceptions,li1996metacognitive} and writing~\cite{zamel1982writing,ramanathan1999individualism, hinkel2003teaching}. 
Hirsch~\cite{hirsch2003reading} argues that \textit{fluency, the breadth of vocabulary,} and \textit{the domain knowledge about the topic} are the three factors that contribute to reading. 
Fluency is the capability to connect current text a person is reading with the preceding text.
Lack of vocabulary creates latency in connecting the previous and current content. 
Lack of domain knowledge can make it  difficult to understand the context, such as contradiction, and eventually create a mental model that is out of the context. 
Note that a lack of one factor can lead to difficulty in other factors.  %

We found additional framing of these challenges in the literature. 
The first challenge is in \textit{text complexity}~\cite{diasti2023implementation,hu2022relative}, where the readers struggle to understand the main gist of the text. 
Text complexity involves not just difficult vocabulary, but also complex grammatical rules and lengthy sentence structures.
All of these factors can lead to misinterpretation or slower reading.%
 Another framing is \textit{reading strategy}~\cite{diasti2023implementation,ramadhianti2023reading,hong14investigating}: inferring the right context, topic, and takeaways from a text block. 
Failure to understand the context, or comprehensive breakdown, is common among EFL readers.

Our approach aims to address these general challenges faced by EFL readers. 
We categorize the challenges into three challenges (vocabulary, grammatical, and comprehension) and addressed them in a unified tool.   
The goal is to provide on-demand support to EFL readers when they face the issues as well as proactively recommend potential issues to improve the self-learning process.
\vspace{-6pt}

\subsection{Intelligent Reading Tools} %
\label{subsec:augmenting-document-reading}

HCI and AI research has a long history of developing intelligent reading tools. The primary focus is to speed up reading~\cite{zhang2023concepteva,lam2024lloom}. For example,  
Marvista~\cite{chen2023marvista} aids users in reading and comprehending text based on the time they want to spend, using AI-generated questions. 
CReBot~\cite{peng2022crebot} interactively asks section-level critical thinking questions, catering to both experienced researchers and novices. Highlighttext~\cite{strobelt16highlighttext} explores effective text highlighting methods, while Living Papers~\cite{heer2023livingpapers} introduces a grammar for augmenting scientific articles. 
SCIM~\cite{fok2023scim} enables researchers to quickly skim papers, and Vogel et al.~\cite{nikhita2024constrainedhigh} show that highlighting a limited number of keywords improves comprehension. 

Several works aim to improve the understanding of charts and tables within documents. 
One approach is adding textual information~\cite{gaba2023langaugevisual,stokes2023balancetextchart} on visualizations. 
Kim et al.~\cite{kim2018doctexttable} link text and tables, while Elastic Documents~\cite{badam2019elasticdocuments} use visualizations to connect them for better summarization. 
Kori~\cite{latif2022kori} provides an interactive interface linking charts and text, and Statslator~\cite{masson2023statslator} visually translates statistical tables for clearer interpretation. Other works focus on improving chart accessibility for readers~\cite{choi19visblind, hoque2023susurrus,kim2023treemapexplain,chundary2024tactualplot}.

Another line of research focuses on using eye movement data to understand and predict confusion~\cite{sims20userconfusion, lalle16predicting} and mind wandering~\cite{dmello16mindwandering} through Machine Learning techniques. 
These studies aim to identify eye movement patterns associated with confusion and use them to predict confusion or mind wandering during reading.

Overall, the above works improve reading experiences for users. However, they do not cater to the needs of EFL readers. The methods also lack focus on the learning aspect, which is critical for EFL readers. Indeed, EFL readers may not only want to read fast but may want to learn about the syntax and semantics. Our work aims to fill that gap.

It is worth noting that researchers have proposed a few tools for learning English. 
For example, ChatPRCS~\cite{wang2024chatprcs} tailors reading comprehension questions to the student's English proficiency. 
VocabEncounter~\cite{arakawa2022vocabencounter} uses recent NLP techniques to integrate vocabulary into the user’s reading material in near real-time. 
CriTrainer~\cite{yuan2023critrainer} helps develop critical paper reading skills through text summarization and template-based questions. Our work extends this line of work by developing NLP methods to proactively identify text that EFL readers may find difficult and then using an LLM to explain the grammar and context of the text. \techname{} also provides on-demand explanations for texts identified by readers as difficult. Finally, the tool uses a verification process to improve the reader trust in the explanation provided, improving the learning experience.

\subsection{Intelligent Writing Tools} %
\label{subsec:english-learning-tools}

Reading and writing are fundamentally related to each other. Learning to write English could potentially improve reading skills.
As such, intelligent writing tools are relevant to our work.  Commercial tools such as Grammarly can help writers improve grammatical and semantical aspects of a written text. There is a new wave of intelligent writing tools powered by LLMs~\cite{lee2022coauthor,kang2022fwyktwyr,karolus2023hardtoread,zyska2023care,wamgsganss2021arguetutor}. Ito et al.~\cite{ito2024airewrite} conduct a study validating AI-assisted drafting of English essays by non-native speakers, developing a tool called Langsmith to refine academic essays.
Figura11y~\cite{singh2024figura11y} supports the writing of scientific alt text. There are efforts to provide customized related work suggestions for scientific articles~\cite{palani2023relatedly, chang2023citesee, kang2023comlittee, hang2022threddy}, often using algorithms based on users' publication history. 
The algorithms mainly deployed in these works are based on users' past records. 
Research search engines like Google Scholar now offer personalized paper recommendations, and there are works ~\cite{lee2024paperweaver} that aim to improve recommendation accuracy with LLMs. Finally, several studies also evaluate the impact of LLMs in human-LLM collaborative writing~\cite{gero2023socialdynamicswriting,li2024concernsinllmwriting, hoque2024hallmark}.
 
This line of research motivated our research. While there are numerous intelligent writing tools (and reading tools), how NLP and LLMs can assist EFL readers remains largely unknown. Our work shows that carefully designed NLP heuristics, along with the LLM can capture a wide range of problems a reader might have and provide proactive and on-demand explanation.

\section{Design Requirements}
\label{sec:03-reading-help-lite}

We adopted an iterative design approach to develop the proposed tool for EFL readers. Before developing our final tool, we created an initial version, \techname{}-Lite, to evaluate the strengths and limitations of our approach. 
Below, we describe our preliminary design requirements, grounded in prior literature.

\begin{itemize}
    \item[\textbf{DR1}] 
    
    \textbf{Proactive and  On-Demand Guidance.} The tool should guide EFL readers on parts that are potentially difficult to parse or understand. Such proactive guidance could help EFL readers focus on textual complexity that they might have overlooked~\cite{diasti2023implementation,hu2022relative} or proceed with the wrong interpretation~\cite{diasti2023implementation,ramadhianti2023reading,hong14investigating}. It is also important that such guidance is provided on-demand, upon the user's request. %

    \item[\textbf{DR2}] \textbf{Explain  vocabulary, grammar, and semantics.} Our tool should provide explanations for improving text comprehension.   The tool should effectively explain unknown vocabulary and  complex grammatical structure. The tool should also explain the core topic and main takeaways of a text segment.  
    
    \item[\textbf{DR3}] \textbf{Drill-down to the Details.} To deliver knowledge and information in a limited amount of computer space, feedback should be arranged hierarchically~\cite{shin2025visualizationary}, where the tool initially provides a high-level summary and the user can drill down to refer to more details~\cite{shin25drillboards}. %

    \item[\textbf{DR4}] \textbf{User-centered adaptive support. } EFL readers may have different levels of profiency in English. The readers' formal education and social exposure to English can also vary. Thus, it is important to provide feedback that is tailored to varying levels of English proficiency. 
    \item[\textbf{DR5}] \textbf{Visual Emphasis.} Visual emphasis techniques~\cite{hoque2024hcaitools} such as word highlights can make the tool effective for two reasons: (1) the emphasis could perceptually guide users to look for salient parts; and (2) it could help readers spend less time in looking for a specific keyword of interest in the text~\cite{strobelt16highlighttext}.

\end{itemize}

\begin{figure*}[t]
    \centering
    \includegraphics[width=\linewidth]{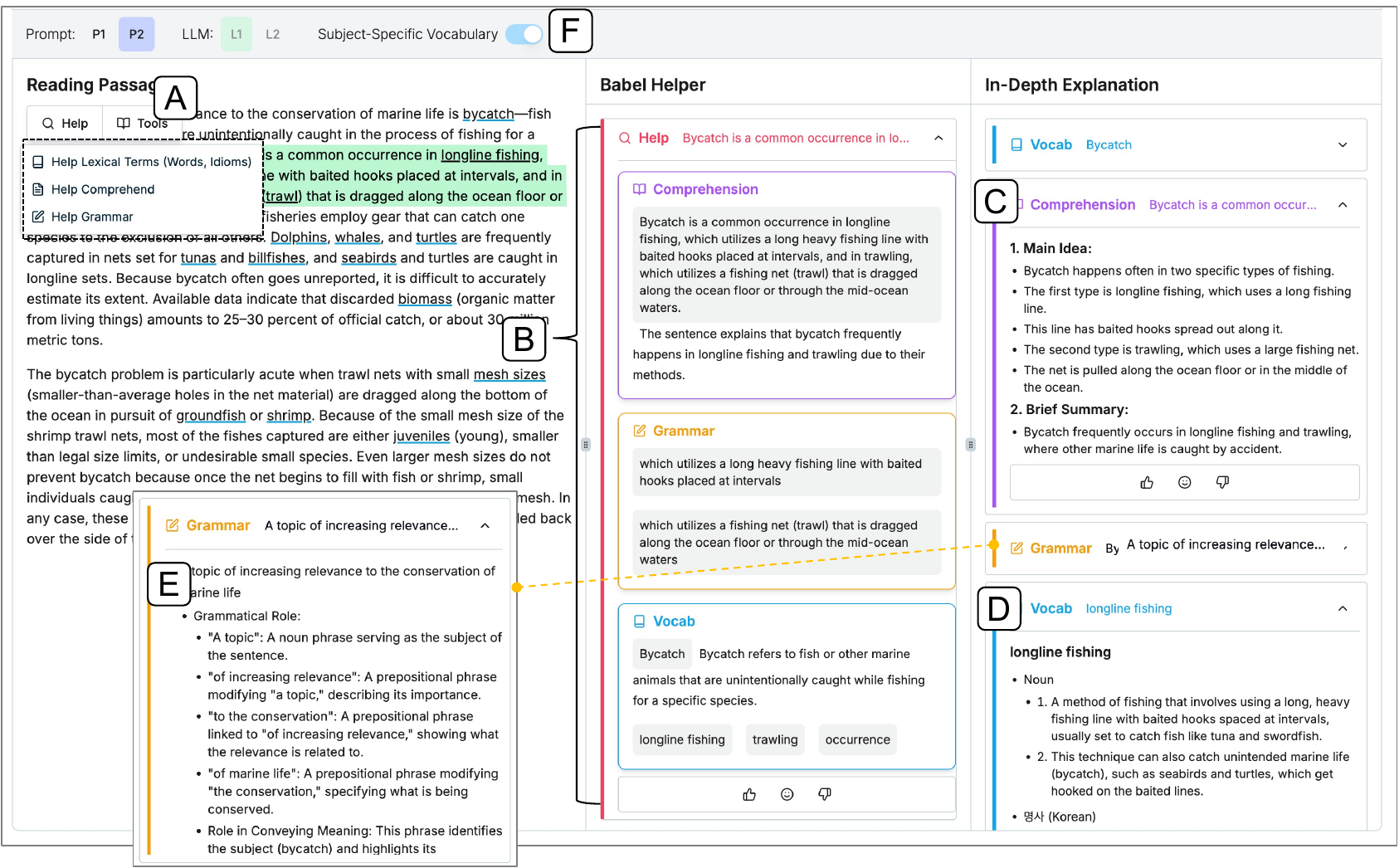}
    \caption{\textbf{\techname{}-lite interface.} (A) A control panel that appears upon selection of a text. The user can choose to get help, or manually look for what they want by pressing the button `tools.' (B) On selecting `help', \techname{}-Lite displays potential challenging issues that the user might be interested in. A user can select any of the issues to see it in a detailed view (C, D, E).
    By toggling the `subject-specific vocabulary,' (F) the user can view subject-specific keywords underlined in the text.  
    }
    \Description{Figure 2 illustrates the Reading.help-Lite interface. Panel (A) shows a control panel that appears upon text selection, with options to either get help or manually activate the ‘tools’ button. When the help option is chosen, panel (B) displays potential challenging issues. Panels (C), (D), and (E) offer detailed views of these issues. In panel (F), toggling the ‘subject-specific vocabulary’ option underlines keywords that are specific to the subject matter.}
    \label{fig:lite_interface}
\end{figure*}

\section{\techname{}-lite}
\label{sec:4-system-design}

Here, we describe the implementation of \techname{}-Lite, our initial attempt to address DR1-DR5. \autoref{fig:lite_interface} shows the interface of \techname{}-Lite. We chose an LLM as the main analytical pipeline for this version because of the ease of access and strong analytical capability. We anticipated that the LLM-powered prototype will allow us to conduct preliminary study with real users and help us design the final version of the tool.

\subsection{Interface Overview}

\subsubsection{Workflow}  
 \techname{}-lite identifies and suggests potential sources of difficulty by analyzing a piece of  text uploaded by a user (DR1). 
The text is analyzed along three dimensions: vocabulary (denoted as `lexical terms'), grammar, and comprehension (DR2). 
The \textit{vocabulary} component identifies unfamiliar keywords or phrases (e.g., idioms and short clauses). 
The \textit{comprehension} component situates the text within the broader context of the passage. 
Finally, the \textit{grammar} component addresses grammatical issues. The three types of issues (i.e., dimensions) are presented in three different types of colored boxes (DR5) (\autoref{fig:lite_interface}B).

\subsubsection{Detailed explanations to users} 
The \textit{vocabulary} module provides the literal definition of a keyword or idiom. 
For words with multiple interpretations (e.g., `take' and `get'), we offer only the context-relevant definition.
A corresponding Korean translation is also supplied for verification (see~\autoref{fig:lite_interface}D).
In the \textit{comprehension} module (see~\autoref{fig:lite_interface}C), we provide the main idea, and the intention of the selected text in relation to the flow of the entire passage. 
The main idea is provided as a bullet point to facilitate the understanding. 
We also provide several examples of sentences that convey the same meaning, but using different wordings and structure.
This is to help users better understand the current context by comparing it with other possible contexts. 
Finally, the \textit{grammar} module (see~\autoref{fig:lite_interface}E) examines sentence structure by segmenting text into phrases. 
For each phrase, the system offers keyword-level explanations that delineate the function of each component. 
The intent behind this is to help users quickly locate issues within a sentence while understanding how individual keywords form the overall phrase.

\subsubsection{Additional Features}
\techname{}-Lite has several other interactive features. 
First, we highlight subject-specific keywords. 
The user can toggle the button next to the `subject-specific vocabulary' label in \autoref{fig:lite_interface}, and the tool underlines subject-specific keywords in blue. 
Hovering over an underlined keyword will show a brief definition of the keyword in a tooltip.
Second, if a user encounters text segments not recommended by the tool but still remains unclear, then they can use the `tools' button in the panel to directly select and analyze the segment by choosing a dimension (vocabulary, grammar, or comprehension). The explanations provided to the user differ bt the proficiency level of the user (DR4) (discussed next).

\subsection{Prompt Engineering}
\label{subsec:prompt-engineering}

We use OpenAI's GPT-4o model for providing explanations to users. 
We iteratively tested various prompting methods to search for the optimal prompt in our setting. 
Effective prompt engineering is challenging as the responses from LLMs are not predictable at times, due to issues such as hallucinaton, verbosity and so on. 
We deploy seven guidelines from  prior work and ChatGPT instructions in designing effective and fast prompts: 

\begin{itemize}[itemsep=0.01cm, topsep=0.15cm, leftmargin=0.4cm]
    \item Assign a role to GPT --- to ensure responses are aligned with a specific expertise (e.g., expert English instructor)
   \item guide LLMs to respond in a specific response template --- to avoid verbose responses.
    \item structure responses clearly and concisely --- to avoid verbosity and maintain precision.
    \item use exact keywords --- to avoid misunderstandings when prompt chaining. 
    \item add global variables (e.g., user proficiency for supporting DR4) in the system prompt --- to  manage a user's specifications equally across various tasks.
    \item add local variables (e,g., tasks) in the user prompt -- to use the same template equally for multiple users. 
\end{itemize}

We explain three main types of prompt template used in the tool. 
The first template is for providing various explanations (e.g., grammar, components, lexical terms) used in \techname{}.
The second template is for computing recommendations about what the user may not know about the user-selected text. 

\subsubsection{Template for providing explanations} Let us define the template for providing explanations as ${Tem}_{ex}.$
Let $[Role]$ be the string that contains the prompt that specifies a role.
We provide the role of an English instructor (e.g., ``\textit{You are an expert English instructor tasked with providing a comprehensive analysis of sentences.}''). 
Let $[Prof_u]$ be the information on the user's English proficiency skills (e.g., `\textit{the user's English proficiency level is `intermediate.}') .
Note that $[Role_I]$ and $[Prof_u]$ are added in the system prompt. 
Let $[Text]$ be the string that contains the text selected by the user. 
$[Text]$ can be a part of the passage, such as `\textit{Because bycatch often goes unreported, it is difficult to accurately estimate its extent.}'
Let $[All-text]$ be the string that contains entire passage provided by the user, and
$[Inst_T]$ be the string that contains of all task instructions needed to conduct the task. 
For example, for providing definitions about a vocabulary, the prompt is \textit{`Provide a comprehensive definition for the given vocabulary word.'}
Finally, let $[Form]$ be the string that contains the exact desirable answer format (e.g., `\textit{Provide the response in the follwing manner: Banana  1. A fruit ... 2. ... }').
To sum up, the template for ${Tem}_{ex}$ is:
$$ {Tem}_{ex} = [Role_{ex}] + [Prof_u] + [Text] + [All-text] + [Inst_{ex}] + [Form_{ex}]. $$

\subsubsection{Template for recommendations (Help)} 
Let us define the template ${Tem}_R$ for providing recommendations in help components (comprehension, grammar, and lexical terms) for the user.
We skip the definitions of $[Role_I]$, $[Prof_u]$, $[All-text]$ and $[Text]$, as they are defined in a similar manner to those in the template above. 
Let $[Inst_R]$ be the string that contains of all task instructions needed to provide recommendations (e.g., `\textit{Please structure your response according to the guidelines provided in the system prompt.}').
Similar to above, $[Form_R]$ is the string that contains the desired exact answer format,
Hence, to sum up, the tamplate for ${Tem}_R$ is: 
$$ {Tem}_R = [Role_I] + [Prof_u] + [Text] + [All-text] + [Inst_R] + [Form_R]. $$

Further details about the prompting and its relevant code can be found in \href{https://osf.io/tf2w8/?view_only=f715f4c297334f55a95410f2c03da596}{here}
\footnote{\url{https://osf.io/tf2w8/?view_only=f715f4c297334f55a95410f2c03da596}}.

\subsection{Implementation details} %
\label{subsec:implementation}

The front end \techname{} is implemented as a full-stack web application using Next.js, a React-based framework for both front-end and back-end development.
In the backend system, the data management and routing are handled entirely within Next.js.
The tool is hosted on Amazon Web Services (AWS) during the user study.

\section{Pilot Study}
\label{sec:5-study-design}

We conducted a pilot study by using \techname{}-lite as a probe. The goal was to evaluate the effectiveness of \techname{}-Lite and identify limitations for improving the tool.
The study was approved by the National Institutional Review Board (IRB) of South Korea. 
We first describe the study procedure, and then provide the results, and takeaways from the study.

\subsection{Study Details}
\label{subsec:study-process}

\noindent\textbf{Participants. }We recruited 15 South Korean participants for this study through the word of mouth approach. 
All participants had a bachelor degree but reported difficulties reading English (i.e., considered themselves as EFL readers). 
Among them 3 participants were English educators. 
Two were high-school teachers whereas the other is a consultant in an educational startup. 
Among the participants, 11 were men, while 4 were women. 
The gender distribution is imbalanced, partly because it was difficult to find adults who would identify as EFL readers as it could be a socially uncomfortable situation for many people.

\smallskip
\noindent\textbf{Task and Materials. }We define a unit task (UT) as a task in which, over the course of 5 minutes, a participant would read a passage ($\sim$200-250 words), highlight any part of the text that they find difficult, and read explanation provided by \techname{}-Lite.
After each session, participants  verbally informed the session administrator about the issues they faced and how the tool helped address the issues.
The passages are derived from TOEFL free reading test samples from ETS~\footnote{https://www.ets.org/pdfs/toefl/toefl-ibt-free-practice-test.pdf}~\footnote{https://www.ets.org/pdfs/toefl/toefl-ibt-reading-practice-sets.pdf}.
We decided to use passages from TOEFL, as it is a popular standardized test for university-level students.

\smallskip
\noindent\textbf{Study Process. } 
Each session consists of an introduction ($\sim$10 minutes), main study ($\sim$35 minutes), and a post-study interview ($\sim$15 minutes). Following consent and the collection of demographic information and relevant background information, participants were guided to  \techname{}-Lite, where the study administrator (the first-author of this paper) demonstrated its features and explained the required tasks.
The main study started with a test to determine the level of English proficiency of the participants. For this test, we asked participants to read a passage for 5 minutes, and answer three questions pertaining to the passage. When a participant answered at most one question incorrectly, we classified them as `proficient'. When they answered more than one question incorrectly, they are categorized as `not proficient.' Based on this categorization, we prompted the LLM to provide easy-to-understand explanations for `not proficient' participants. After that, participants completed four unit tasks. 

During the post-interview study, we inquired about the effectiveness of the tool and the components and their general satisfaction with the tool. We also asked about limitations and failures of the tool. Each session lasted around 60 minutes.
We provided a compensation of around KRW 15,000.- (around US\$12.00) for the compensation.
This rate is based on the hourly minimum rate of KRW 9,860/hr.

\subsection{Results}
\label{subsec:pilot-results}

Overall, the feedback from the participants were positive. They saw the value of such a tool for personal growth and improvement. We discuss the major findings below.

\begin{figure}[tbh]
    \centering
    \includegraphics[width=1.0\linewidth]{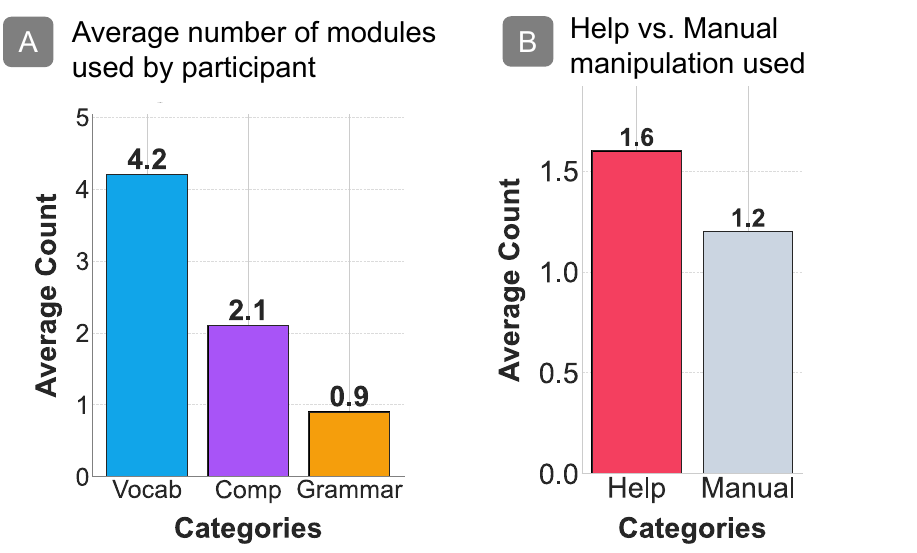}
    \caption{\techname{}-Lite usage during pilot study. 
    (A) shows the average number of modules (vocabulary, comprehension, and grammar) used by participants while conducting a unit task. 
    (B) compares the number of times `help' and manual manipulation, or `tools' are accessed while conducting a unit task. }
    \Description{Figure 3 presents the usage analysis during a pilot study of Reading.help-Lite. Panel (A) shows the average number of modules used by participants—covering vocabulary, comprehension, and grammar—while conducting a unit task. Panel (B) compares the frequency of accessing ‘help’ versus manual activation of the ‘tools’ button, illustrating usage patterns during the experiment.}
    \label{fig:userstudy-chart}
\end{figure}

\smallskip
\noindent\textbf{Usage patterns. } We noticed that participants sought the most help with vocabulary, followed by grammatical structure. 
On average, participants sought 4.2 explanations for vocabulary, 2.1 for comprehension, and 0.9 for grammar, per each passage within a document (see~\autoref{fig:userstudy-chart}A). We find that participants used the `help' button 1.6 times to identify all issues for a selected text. In contrast, participants  sought specific explanation (that is either comprehension, grammar, or lexical terms) for a selected text 1.2 times (~\autoref{fig:userstudy-chart}B). 
While most participants found the feedback accessible, some users with lower TOEFL proficiency (4/15) reported that the responses were overly verbose and difficult to comprehend.

\smallskip
\noindent\textbf{Post-study interview. }To begin with, All participants reported that LLM recommendations were beneficial, though they did not expect them to be completely error-free.
Second, participants expressed a strong preference for the comprehension module (9/15) and the vocabulary module (8/15) among the deployed components. 
The comprehension module was utilized to decode lengthy, complex sentences. 
As one participant explained, \textit{``The comprehension component explains difficult parts by showing the sentence's intention and considering the passage's overall context. This helps people spot their misinterpretations.''}

\subsection{Lessons Learned}
\label{subsec:lessons-learned}

We identified several challenges that must be addressed to make the tool more effective. Following limitations informed the revised tool in \S~\ref{sec:reading-help}.

\begin{itemize}[itemsep=0.01cm, topsep=0.15cm, leftmargin=0.4cm]

    \item \textbf{Lack of trust and validation.} There is a consensus among participants that the responses of LLMs are insightful, but it is difficult to trust the responses. Participants were not confident in accepting the responses of the LLMs in some cases.  Several participants queried about how the system is determining words that might be challenging. 
    \item \textbf{Lack of adequate support for comprehension. } While \techname{}-Lite provided comprehension support for a selected text, we did not provide comprehension support for the whole document. 
    The current selection based approach assumes that users will recognize what they do not know. 
    However, we have observed that users may be uncertain about where to look for additional guidance. 
    This issue is also echoed in the literature~\cite{ramadhianti2023reading} that lack of effective English reading strategies also impedes EFL reading.
    \item \textbf{Verbosity of the grammar explanation.} the grammar component was widely regarded as the least useful, with 12/15
participants reporting minimal benefit. 
Participants were overwhelmed by its verbosity. Several participants suggested that this component could be really useful if the explanation is brief and concise.

    \item \textbf{Adapting to varying proficiency levels. } In our study, participants with lower TOEFL scores in three-question test received simplified responses. 
    Reactions varied: some found the suggestions acceptable, while others struggled to comprehend them, indicating that the responses could be made more accessible for EFL readers in general. %
    \item \textbf{Usability issues. } We also found some usability issues regarding \techname{}-Lite, such as the confusion with the different panels in the tool and lack of explanations for the colors used in the interface. 
\end{itemize}

\section{\techname{}: Updated Version }
\label{sec:reading-help}

\begin{figure*}[t]
    \centering
    \includegraphics[width=\linewidth]{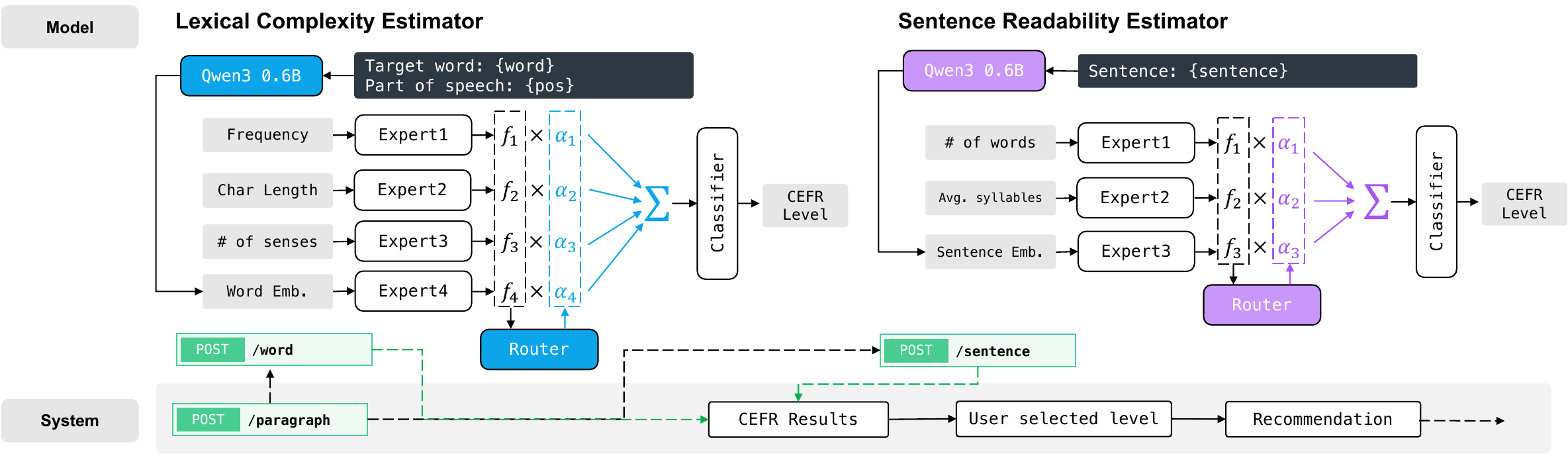}
    \caption{Overview of our interpretable CEFR prediction models. \textbf{Left:} Lexical Complexity Estimator (word-level). \textbf{Right:} Sentence Readability Estimator (sentence-level). Each feature is processed by a small expert network \(f_i\); a data-dependent router predicts weights \(\alpha\) that gate the experts and form a weighted sum, which a classifier maps to a CEFR level (A1-C2). Word features: frequency, character length, number of senses, and a contextual embedding from a LoRA-adapted Qwen3 0.6B~\cite{qwen3}. Sentence features: number of words, average syllables per word, and a contextual embedding from the same model. The gray callouts show the prompts used to obtain contextual embeddings. The learned \(\alpha\) weights provide instance-specific, explanations of which features drove each prediction. The word- and sentence-level CEFR results are combined, then filtered to include only items at or above the user-selected CEFR level, and presented as the final recommendations.}
    \Description{Figure 4 is a two-panel block diagram of interpretable CEFR prediction. The left panel (Lexical Complexity Estimator, word level, blue) takes four word features—frequency, character length, number of senses, and a contextual embedding from a LoRA-adapted Qwen3-0.6B—and feeds each into a small expert; a router assigns weights $\alpha$ that gate the experts, a Σ node forms a weighted sum, and a classifier outputs a CEFR label (A1–C2). The right panel (Sentence Readability Estimator, sentence level, purple) mirrors this pipeline with sentence features—number of words, average syllables per word, and a contextual sentence embedding—using three experts, router weights $\alpha$, a Σ node, and a classifier to produce a CEFR label. Gray callouts show example prompts used to obtain the contextual embeddings. A system row (green) illustrates API flow: a paragraph is posted to the backend (`POST /paragraph`), which yields word-level (`POST /word`) and sentence-level (`POST /sentence`) analyses, producing CEFR results; solid arrows indicate forward data flow and dashed lines depict service connections. The word- and sentence-level outputs are combined and filtered to retain only items at or above a user-selected CEFR level, which are then presented as recommendations; the learned $\alpha$ weights provide instance-specific explanations of which features most influenced each prediction.}
    \label{fig:cefr-model}
\end{figure*}

We develop \techname{} by addressing the issues we found in \techname{}-Lite (see \S\ref{subsec:lessons-learned}). 
In this section, we describe the revised system with the key modifications. 

\subsection{Interpretable CEFR Prediction Models}
\label{subsec:interpretable-model}
One of the main concerns from the pilot study was that \techname{}-lite was not transparent. Participants were hesitant to accept the recommendation without a clear decision-making process. Thus, we decided to develop an interpretable model for recommending challenging words and sentences. Another concern was that participants could not weigh the severity of a recommendation (that is, how complex the word or sentence is) without referring to a baseline. Thus, we decided that the recommendation should align with a standard practice. We adopted the Common European Framework of Reference (CEFR)~\cite{council2001common} for Languages, a standardized system of six levels (A1, A2, B1, B2, C1, C2) to measure language proficiency. This standard is typically used to prepare learning materials that match the six levels. A1 and A2 indicate basic users, B1 and B2 indicate independent users, and C1 and C2 proficient users. Prior research shows that CEFR is an effective measure to calculate lexical complexity. We propose two interpretable neural models for CEFR level prediction: a Lexical Complexity Estimator and a Sentence Readability Estimator.

\smallskip
\subsubsection{Lexical Complexity Estimator}
Suppose a document $D$ contain $S$ sentences and $W$ words. Denote the $i$-th sentence by $s_i$ and the $j$-th word by $w_j$. The Lexical Complexity Estimator is defined as the following function $f$:
$$f: X \rightarrow \{0,1\}^C$$
where $X$ is the input feature space and $C$ denotes the number of CEFR classes. Our Lexical Complexity Estimator is designed to classify individual words into CEFR levels based on four linguistically meaningful features $x_i$ inspired by prior work~\cite{kelious2024complex,aleksandrova2023cefr,fujinuma2021semi}: 1) frequency or rarity of the word in English ($freq(w_j)$); 2) length or character count of the word ($|w_j|$); 3) number of senses or polysemy ($poly(w_j)$); and 4) the word embedding of $w_j$. To compute the word embedding, we use the Qwen3-0.6B~\cite{qwen3} model. We provide the word and its part of speech to Qwen3 model in a prompt and take the mean of the final hidden states as the embedding. The model architecture consists of four main components (\autoref{fig:cefr-model}A). First, \textbf{feature transformation networks} process each of the four features through dedicated transformation networks, acting as specialized experts:
\begin{align*}
f_i &= \text{Expert}_i(x_i) \\
    &= \text{LayerNorm}(\text{GELU}(\text{Linear}(\text{LayerNorm}(\text{Linear}(x_i)))))
\end{align*}

Here, each expert network transforms a scalar feature into a 512-dimensional representation via two linear layers with LayerNorm~\cite{ba2016layernormalization} and GELU~\cite{hendrycks2023gaussianerrorlinearunits} activations. 

Second, \textbf{dynamic alpha prediction} uses a data-dependent gating mechanism. Instead of fixed feature weights, we use an alpha predictor network that determines the importance of each feature based on the current input:
$$
\alpha = \text{softmax}(\text{AlphaPredictor}(\text{concat}(f_1, f_2, f_3, f_4)))
$$

Third, \textbf{feature fusion} combines the weighted features through element-wise multiplication and summation:
$$
h = \sum_{i=1}^{4} (\alpha_i \times f_i)
$$

Finally, the fused representation is passed through a final classifier network to produce CEFR class logits.
\begin{align}
z &= \text{Classifier}(h),\quad z \in \mathbb{R}^{K} \\
\hat{y} &= argmax_{k \in \{1,\dots,C\}} z_{k}
\end{align}

The final \textbf{classification layer} is a small feed-forward network with two linear layers. It first reduces the feature dimension by half, then normalizes the activations with LayerNorm, and applies the GELU activation function, similar to feature transformation networks. Finally, another linear layer maps the features to the number of output classes.

\begin{table*}[t]
\centering
\caption{CEFR prediction results (\%). Best per column (within each category) in \textbf{bold}.}
\Description{This table compares CEFR prediction performance (F1 by level A1–C2 and overall metrics) for two categories—Word CEFR and Sentence CEFR—between GPT-4o and our model. For Word CEFR, our model outperforms GPT-4o at every level, reporting F1 scores of 81.2 (A1), 57.9 (A2), 53.0 (B1), 68.2 (B2), 63.3 (C1), and 67.5 (C2), versus 68.5, 40.4, 51.8, 59.3, 41.2, and 49.5, and achieves higher overall results (Avg. F1 65.2 vs. 51.8; Accuracy 64.8 vs. 52.9; Grouped Accuracy 71.1 vs. 62.3). For Sentence CEFR, our model leads strongly from A1 through C1 with F1 scores of 70.6, 86.3, 90.0, 87.6, and 86.8 compared with 50.0, 59.5, 62.0, 71.1, and 65.8, but trails at C2 (28.6 vs. 49.0 for GPT-4o); nonetheless, it delivers substantially better overall results (Avg. F1 75.0 vs. 59.6; Accuracy 87.4 vs. 64.0; Grouped Accuracy 89.7 vs. 69.9). Boldface entries indicate the best score in each column within its category.}

\begin{tabular}{llccccccccc}
\toprule
 & & \multicolumn{6}{c}{Levels F1 (\%)} & \multicolumn{3}{c}{Overall (\%)} \\
\cmidrule(lr){3-8}\cmidrule(lr){9-11}
Category & Model & A1 & A2 & B1 & B2 & C1 & C2 & Avg. F1 & Acc. & \makecell[c]{Grouped Acc.} \\
\midrule
\multirow{2}{*}{\makecell[l]{Word CEFR}}
 & GPT-4o & 68.5 & 40.4 & 51.8 & 59.3 & 41.2 & 49.5 & 51.8 & 52.9 & 62.3 \\
 & Ours             & \textbf{81.2} & \textbf{57.9} & \textbf{53.0} & \textbf{68.2} & \textbf{63.3} & \textbf{67.5} & \textbf{65.2} & \textbf{64.8} & \textbf{71.1} \\
\midrule
\multirow{2}{*}{\makecell[l]{Sentence CEFR}}
 & GPT-4o & 50.0 & 59.5 & 62.0 & 71.1 & 65.8 & \textbf{49.0} & 59.6 & 64.0 & 69.9 \\
 & Ours             & \textbf{70.6} & \textbf{86.3} & \textbf{90.0} & \textbf{87.6} & \textbf{86.8} & 28.6 & \textbf{75.0} & \textbf{87.4} & \textbf{89.7} \\
\bottomrule
\end{tabular}
\label{tab:cefr_results}
\end{table*}

\subsubsection{Sentence Readability Estimator}
The estimator follows the same functional form for a sentence $s_i$:
$$g: Y \rightarrow \{0,1\}^C$$
where $Y$ is the sentence-level feature space. This estimator follows the same design principles but is based on three features inspired by prior work~\cite{arase2022cefr,naous2024readme++}: 1) number of words ($|s_i|$); 2) average syllables per word ($syl(s_i)$); and 3) the sentence embedding of $s_i$. The sentence embedding likewise uses the Qwen3 model. The architecture uses three expert networks and an alpha predictor, with the same fusion and classification approach.

\smallskip
\subsubsection{Interpretability}
A key advantage of our approach is interpretability through dynamic gating. The alpha weights $\alpha$ provide instance-specific explanations of which features contributed to each prediction. This enables feature-level interpretability (each weight corresponds to a linguistically meaningful feature) and instance-specific explanations (the gating is data-dependent).

\smallskip
\subsubsection{Model Training}
Both models are trained end-to-end using the PyTorch library for CEFR classification. The Qwen3-based embedding model is trained with a LoRA adaptation. In multi-label scenarios where a word or sentence may belong to multiple CEFR levels simultaneously, we treat each CEFR level as an independent binary classification task and use multi-label cross-entropy loss (\texttt{BCEWithLogitsLoss}). The training configuration is as follows: batch size 64, learning rate 2e-4, 10 epochs using the AdamW optimizer~\cite{loshchilov2017decoupled} on a single NVIDIA A100 GPU.

\subsubsection{Datasets}

For the sentence dataset, we merged the SCoRE and WikiAuto datasets provided with CEFR-SP~\cite{arase2022cefr}. We also used CEFR-SP's predefined split to construct the train and test sets. For the word dataset, following previous work~\cite{fujinuma2021semi}, we used CEFR-J~\cite{uchida2018assigning} and the Octanove Vocabulary Profile~\cite{fujinuma2021semi}, and additionally incorporated the Oxford 5000~\cite{oxford5000} dataset. In CEFR-J, the word list is based on major English textbooks used in China, Korea, and Taiwan. To reduce label ambiguity, we exclude cases where the same (word, POS) appears in two or more sources but differs by $\geq 2$ CEFR levels. We then split the data into training and test sets with a 90\%/10\% ratio.

\begin{figure*}[ht]
    \centering
    \includegraphics[width=\textwidth]{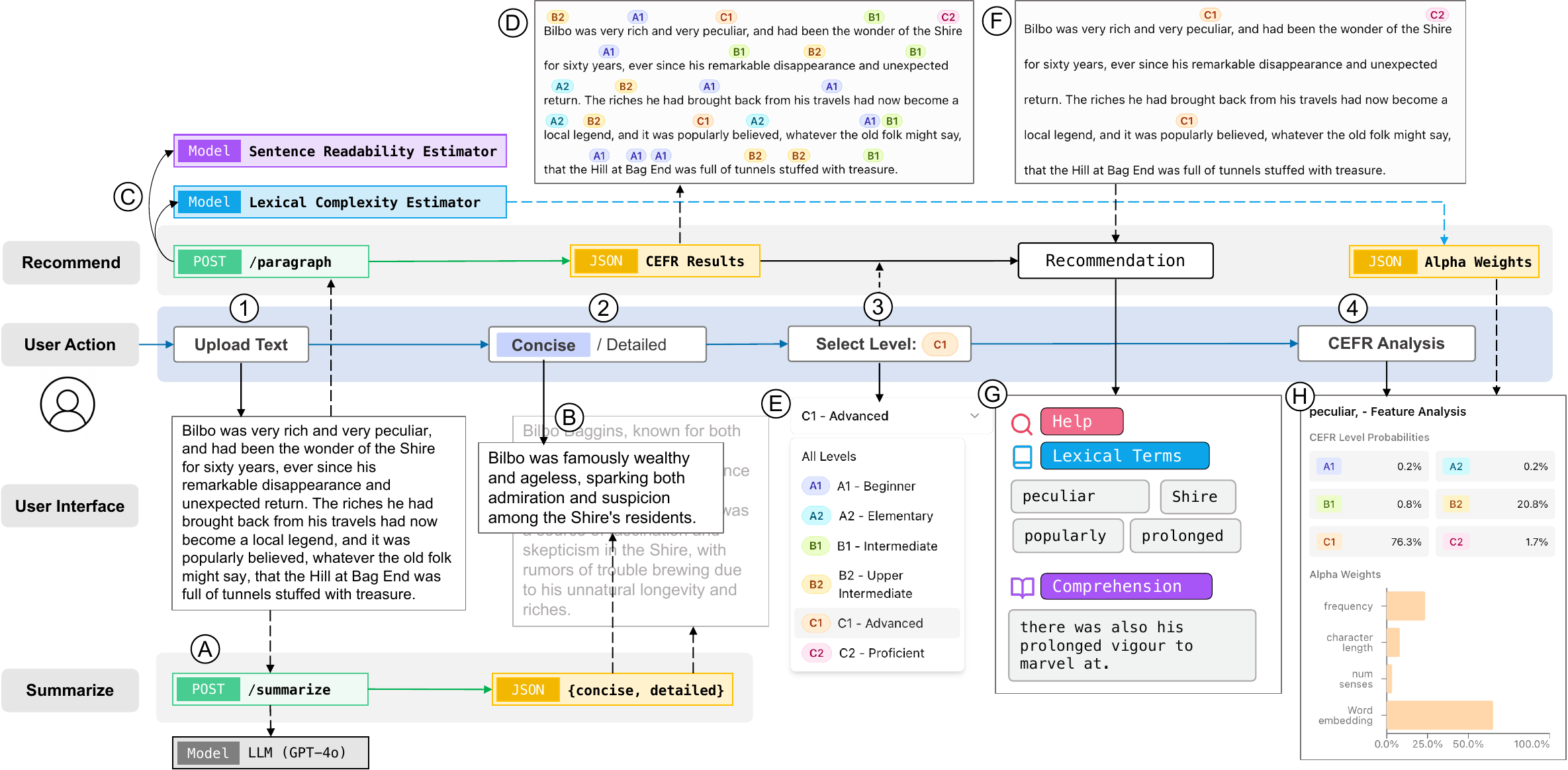}
    \caption{\textbf{User Actions and System Pipeline for Summarization and Proactive Recommendations.}
    Users first upload a document (\textcircled{1}); then choose a summary style, \emph{Concise} or \emph{Detailed} (\textcircled{2}); set a target CEFR level (e.g., B2 or C1) (\textcircled{3}); and run the CEFR analysis (\textcircled{4}). 
    In the system, the backend endpoint \texttt{POST /summarize} generates paragraph-aligned summaries (A), which are displayed in a fixed left sidebar at the selected detail level (B). 
    Interpretable CEFR predictors, the Sentence Readability Estimator (SRE) and the Lexical Complexity Estimator (LCE), perform batched inference and serve results (C). 
    The passage view is annotated with CEFR difficulty labels at the token and sentence levels (D), and proactive recommendations surface words and sentences that exceed the user’s threshold (E, F). 
    A help panel lists recommended items that are likely to be difficult for the user (G). 
    When the user requests a CEFR decision, an interpretable feature analysis presents level probabilities and compact feature-contribution bars (H).}
    \Description{Flow diagram of the user flow and system pipeline: (1) the user uploads a document, (2) selects a summary style (Concise or Detailed), (3) chooses a target CEFR level (e.g., B2 or C1), and (4) runs CEFR analysis; the backend \texttt{POST /summarize} returns paragraph-aligned summaries that appear in a fixed left sidebar (A,B). Two interpretable predictors—the Sentence Readability Estimator and the Lexical Complexity Estimator—run batched inference (C) to annotate the passage with color-coded CEFR tags A1–C2 at token and sentence levels (D). A level selector (E) controls proactive recommendations (F) that surface words and sentences whose predicted difficulty meets or exceeds the chosen level. A help panel (G) lists recommended lexical items and comprehension notes as selectable chips with brief explanations. An interpretable analysis card (H) shows CEFR level probabilities and compact feature-contribution bars (alpha weights) for features such as frequency, character length, number of senses, and contextual embedding; solid arrows indicate user flow and dashed connectors indicate JSON outputs (CEFR Results, Alpha Weights).}
    \label{fig:user_action_system_pipeline}
\end{figure*}

\subsubsection{Model Evaluation}

Table~\ref{tab:cefr_results} reports F1 by level, overall accuracy, and grouped accuracy (A$^\ast$={A1,A2}, B1, B2, C$^\ast$={C1,C2}). We compare interpretable neural models with GPT-4o on CEFR classification tasks at both the word and sentence levels. The dataset may have up to two levels (either because there are multiple annotators or due to the process of merging datasets), and both levels were regarded as equally reliable. Therefore, during testing, a prediction was counted as correct if it matched either of the annotated labels. The prompts provided to GPT-4o were as follows.

\promptbox{Word–level prediction}{%
Task: Predict the CEFR level (A1, A2, B1, B2, C1, C2) of the target word.

Target word: \{word\}

Part of speech: \{pos\}

CEFR Level:
}
\promptbox{Sentence–level prediction}{%
Task: Predict the CEFR level (A1, A2, B1, B2, C1, C2) of the given sentence.

Sentence: \{sentence\}

CEFR Level:
}

For word-level CEFR prediction, our Lexical Complexity Estimator significantly outperforms GPT-4o on all metrics. The overall average F1 of 65.2\% represents a substantial improvement over GPT-4o’s 51.8\%. Likewise, our model achieves higher overall accuracy (64.8\% vs.\ 52.9\%) and grouped accuracy (71.1\% vs.\ 62.3\%). These results suggest that our feature-aware architecture captures frequency, length, and sense (polysemy) based lexical difficulty more reliably than a general purpose LLM. This aligns with prior work showing that, in complex word identification, ChatGPT performs poorly on calibrated scoring while task-specific models achieve stronger performance~\cite{kelious2024complex}.

For sentence-level CEFR prediction, our Sentence Readability Estimator achieves higher F1 scores than GPT-4o on five of the six CEFR levels. GPT-4o is superior only at C2, yet our model still obtains a markedly higher average F1 (75.0\% vs.\ 59.6\%). Our model also outperforms on overall accuracy (87.4\% vs.\ 64.0\%) and grouped accuracy (89.7\% vs.\ 69.9\%). These findings are consistent with prior research showing that fine-tuned models outperform few-shot, prompt-based LLMs in readability assessment~\cite{naous2024readme++}.

\smallskip
\subsubsection{Inference efficiency.}
To compare runtime speed, we measured the average inference time per example on the test set with batch size~1. Our models were executed on a single NVIDIA~A100 GPU, and GPT-4o latencies were measured as end-to-end API round-trip times from the same client machine. The proposed models are substantially faster than GPT-4o on both tasks: for word-level prediction, $0.05$\,s vs.\ $0.611$\,s ($\approx\!12\times$ speedup), and for sentence-level prediction, $0.052$\,s vs.\ $0.592$\,s ($\approx\!11\times$ speedup). This advantage stems from the compact, feature-aware architecture, and in practical deployments latency can further decrease with batch processing.

In summary, the proposed interpretable neural models consistently surpass GPT-4o across both tasks. These results confirm that our specialized, interpretable models are advantageous for CEFR-level classification and are well-suited for deployment in our system.

\subsection{Automated Summary}
\noindent To address DR2 (improving text comprehension), the updated system automatically produces short, paragraph-aligned summaries as soon as a document is uploaded (\autoref{fig:user_action_system_pipeline}A). These summaries foreground topic sentences and key claims so that EFL readers can quickly grasp the main idea before reading the paragraph in detail. Each summary is anchored to its source paragraph in the main text, minimizing context switching and preserving local coherence during skimming and review. A summary toggle offers two settings (\autoref{fig:user_action_system_pipeline}\textcircled{2}): \emph{concise} shows a short cue focused on the topic sentence, and \emph{detailed} provides a detailed information. This lets readers choose how much help they want while keeping each summary anchored to its paragraph, reducing context switching and supporting the reading strategies.

\subsection{Validation}

While the CEFR model will improve interpretability of the identification task, we are still relying on LLMs to generate the explanation for grammar because of its generational capabilities. To improve the reliability of the explanation,
 we added a validation mechanism along with our interpretable models.

\begin{figure*}[t]
    \centering
    \includegraphics[width=\textwidth]{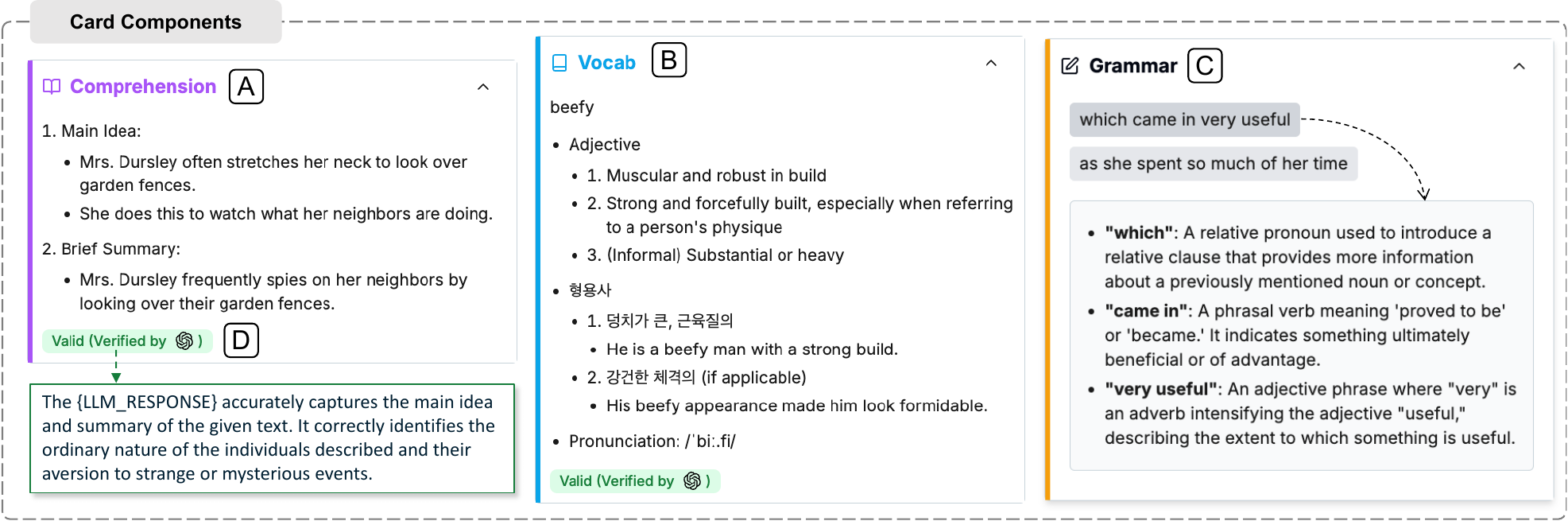}
    \caption{\textbf{Detailed explanation of components to help EFL academic reading.}
    We provide three components to help EFL researchers in academic reading.
    They are: (A) vocabulary (lexical terms), (B) comprehension, and (C) grammar. 
    In (A), we present the definition of the selected keyword, or the idiom, usage, as well as its meaning in Korean. 
    In (B), we help users understand the text by analyzing the main idea of the user-selected text with respect to the entire context. 
    In condition (C), we provide grammatical explanations about the user-selected text at the phrase level.
    }
    \Description{This figure details three key components designed to assist EFL researchers in academic reading: (A) Vocabulary, which provides definitions, usage examples, and Korean translations for selected keywords; (B) Comprehension, which analyzes the main idea of a selected text segment in relation to its overall context; and (C) Grammar, which offers phrase-level grammatical explanations for user-selected text.}
    \label{fig:detailed-horizontal}
\end{figure*}

After the primary LLM generates an explanation (lexical term, comprehension note, or grammar analysis), a second LLM (i.e., validator) checks whether the response (i) is grounded in the selected span or paragraph, (ii) matches the requested assistance type, and (iii) is linguistically correct and non-contradictory. The validator returns a binary decision (\textsc{valid}/\textsc{invalid}) and a brief rationale. Validated items are visually indicated in the UI (\autoref{fig:detailed-horizontal}D). This dual-LLM scheme does not eliminate all errors but makes trust cues explicit to users with modest added latency. See \autoref{tab:validation_accuracy} for results.

\subsection{UI and Usability Improvements}

\smallskip
To enhance usability, we refined the interface with several improvements. We made two changes to make the interface adaptable to different proficiency levels. First, readers can set a target difficulty level from a dropdown with different \emph{recommended CEFR levels} (\autoref{fig:user_action_system_pipeline}E). Second, we introduced a toggle button that allows users to control the level of details in the summaries (either detailed or concise).

The updated interface is more proactive in providing guidance. 
Uploaded documents are automatically processed to produce short, paragraph-aligned summaries displayed in a fixed left sidebar (\autoref{fig:user_action_system_pipeline}B). 
In addition, the system recommends potentially difficult words and sentences (\autoref{fig:user_action_system_pipeline}G). The recommended words and sentences are highlighted with three colors---yellow for grammar, blue for vocabulary, and purple for comprehension. Interacting with a highlighted item reveals its CEFR label, and hovering over it presents a compact attribution chart that indicates which features contributed the most to the prediction of difficulty level (\autoref{fig:user_action_system_pipeline}H).

To minimize the verbosity of the grammar module, we prompted the LLM to make the explanation brief.
Finally, validation by the second LLM is visually indicated in the UI: green icons mark responses confirmed by the second LLM, and hovering over the icon reveals the rationale behind the decision (\autoref{fig:detailed-horizontal}D).

\smallskip
\noindent Original features from \techname{}-Lite are preserved: users may still highlight spans of interest and obtain detailed explanations on demand (\autoref{fig:lite_interface}D).

\begin{figure*}[t]
    \centering
    \includegraphics[width=\textwidth]{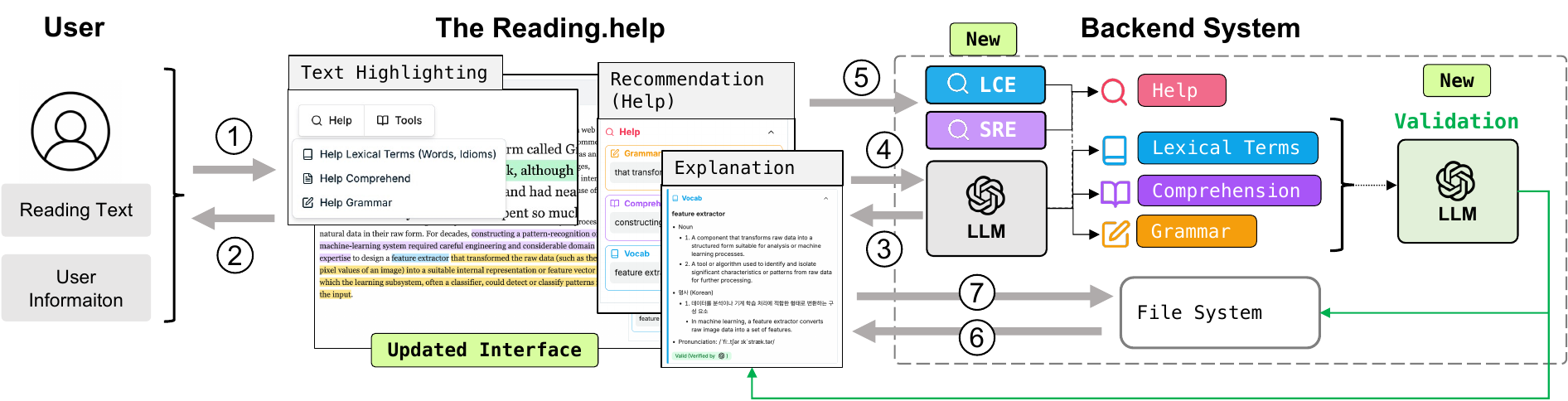}
    \caption{\textbf{System overview.}
    The mechanism of our tool is composed of an interface (the \techname{}) and a back-end system that provides and stores responses. 
    \textcircled{1} The user delivers their demographic information, as well as the part of text that they do not know to the back-end system via the \techname{}.
    Then, \techname{} receives the input and delivers it to the LLMs \textcircled{4}. 
    Afterwards, LLMs provide the responses to the \techname{} \textcircled{3}, and they are delivered to the user \textcircled{2}.
    \textcircled{5} When a document is uploaded, the backend also runs the Lexical Complexity Estimator (LCE) for words and the Sentence Readability Estimator (SRE) for sentences to estimate CEFR levels across the document.
    When a new input is uploaded and new response comes, then the information existing in the interface is saved to the file system in the back-end \textcircled{7}, and can be called back by the interface upon the user's request \textcircled{6}.
    Note that the new validation mechanism is added in \techname{}.
    The validation takes place prior to sending the data to the \techname{} interface from the backend server. 
    }
    \Description{This figure provides an overview of the system architecture for Reading.help. It depicts a two-part mechanism: the user interface and the back-end system. The process begins with the user submitting demographic information and unknown text via the interface. The input is then sent to large language models (LLMs) that generate responses. These responses are delivered back to the interface. New inputs and responses are saved to a file system for later recall, and a validation step (integrated into Reading.help) checks the data before it is shown to the user. Numbered flow indicators trace the steps from input submission to data storage.}
    \label{fig:system-overview-new}
\end{figure*}

\subsection{System Architecture and Implementation}
\label{subsec:system-structure}
\label{subsec:readinghelp-final}

The system comprises two components: a web-based interface for reading assistance (\techname{}) and a backend that consists of CEFR model and LLMs. Here, we describe the system architecture (\autoref{fig:system-overview-new}) and analysis pipeline.

\noindent\textbf{Document preprocessing.}
When a document is uploaded, the backend runs the Lexical Complexity Estimator (LCE) for words and the Sentence Readability Estimator (SRE) for sentences to estimate CEFR levels across the document. 

\noindent\textbf{CEFR Predictors and serving.}
The backend of \techname{} integrates interpretable models with a validation mechanism to deliver reliable assistance at scale.
Both interpretable models are implemented in PyTorch and exposed to the interface via a FastAPI~\footnote{\url{https://fastapi.tiangolo.com/}} server.
To reduce latency, the pipeline executes \emph{batched} inference over tokens and sentences rather than per-item calls.

\noindent\textbf{Processing workflow.}
The pipeline has five stages:
(1) segment uploaded documents into paragraphs;
(2) split into sentences and tokens;
(3) run batched LCE and SRE inference to obtain CEFR estimates;
(4) reassemble predictions into the original text structure; and
(5) filter predictions according to the user's target CEFR level (or history-inferred default),
ensuring that only items above the chosen threshold are surfaced in the interface.

\noindent\textbf{Response generation and validation.}
When the primary LLM generates a candidate explanation for a selected span, a secondary validator LLM evaluates it using a structured prompt containing \{context\_paragraph, selected\_span, component\_type, explanation\}.
The validator returns a JSON object \{\texttt{valid}, \texttt{rationale}\}, which the decision logic parses to annotate the item.
The interface renders this annotation by attaching an indicator and, when available, surfacing the rationale (see \autoref{fig:system-overview-new}).
Although not perfect, this layered approach aligns with established validation techniques in AI systems~\cite{gou2024criticlargelanguagemodels}.

\section{Evaluation}
\label{sec:evaluation}

Our evaluation focuses on three aspects: (1) understanding how \techname{} helps EFL readers, (2) the effectiveness of recommendations based on CEFR models, and (3) the robustness of LLM validation.
For this, we again conducted a user study with 5 participants, which is described in \autoref{tab:participant}.
After detailing the study procedure, we report our results from the study.

\subsection{Study on EFL readers}

We conducted a user study with five EFL readers, all at or above the undergraduate level, who received their entire education, including English instruction, in South Korea.
All participants reported Korean as their most proficient language for both speaking and writing, and none majored in English-related fields.
The study consisted of two phases, with the main session lasting approximately 50 minutes, followed by a 10-minute post-interview.

\paragraph{Phase 1.}
In Phase 1, participants were provided with five separate passages as plain text, without any reading support tools. 
Each passage consisted of four paragraphs and ranged from 340 to 370 words. 
Passages were selected from the of two novels, two \textit{Economist} articles, and one deep learning research paper to evaluate our tool's applicability across  various text types.
Participants were asked to read all the passages for 30 minutes and highlight any sections they found difficult.

\paragraph{Phase 2.}
In Phase 2, participants were introduced to the web-based \techname{} tool. 
After a brief five-minute tutorial, participants revisited the five passages using \techname{}. 
They were asked to interpret text segments they had previously marked as unclear and use \techname{} to resolve the comprehension issue. 
Each participant then detailed the issue, assessed whether the tool provided effective support, and explained how it addressed the problem. 
The session ended after 20 minutes.
Then, we conducted a brief post-interview to gather additional insights on their experiences and \techname{}'s effectiveness.
Their usages of modules in \techname{} are summarized in \autoref{fig:overview-chart}.

\begin{table}[t!]
    \centering
    \caption{\textbf{User study demographics.}
    A total of 5 participants with backgrounds in medicine, computer science, and engineering took part in the study. All participants were above the age of 25 and held at least a bachelor’s degree in their respective fields. They were all Korean nationals who regularly engage with English academic texts as part of their professional or academic work. We present the participants’ gender, age, education (Edu.), and job title. }
    \scalebox{0.99}{
        \begin{tabular}{llllll}
        \toprule
        \textbf{ID} &
        \textbf{Gender} &
        \textbf{Age} &
        \textbf{Edu.} &
        \textbf{Job Title} &
        \\
        \midrule
        P1 & Male & 30 & M.D. & Medical Doctor \\
        \rowcolor{gray!20}
        P2 & Female & 29 & M.D. & Medical Doctor \\
        P3 & Female & 27 & M.S. & Ph.D.\ student in CS \\
        \rowcolor{gray!20}
        P4 & Female & 26 & M.S. & Ph.D.\ student in CS \\
        P5 & Male & 29 & B.S. & Software engineer \\
        \bottomrule
    \end{tabular}
    }
    \Description{This table presents the demographic details of the 5 study participants. The table lists each participant’s gender, age, education level (all hold at least a bachelor’s degree), and job title. All participants are Korean nationals engaged in English academic reading in fields such as medicine, computer science, and engineering.}
    \label{tab:participant}
\end{table}

\subsection{Results}
\label{subsec:results-readinghelp}
We report our findings in three parts. 
First, we describe general usage patterns of participants. 
Next we evaluate the effectiveness of the recommendation of \techname{}. Finally, we assess the validity of the LLM.

\begin{figure}[tbh]
    \centering
    \includegraphics[width=0.9\columnwidth]{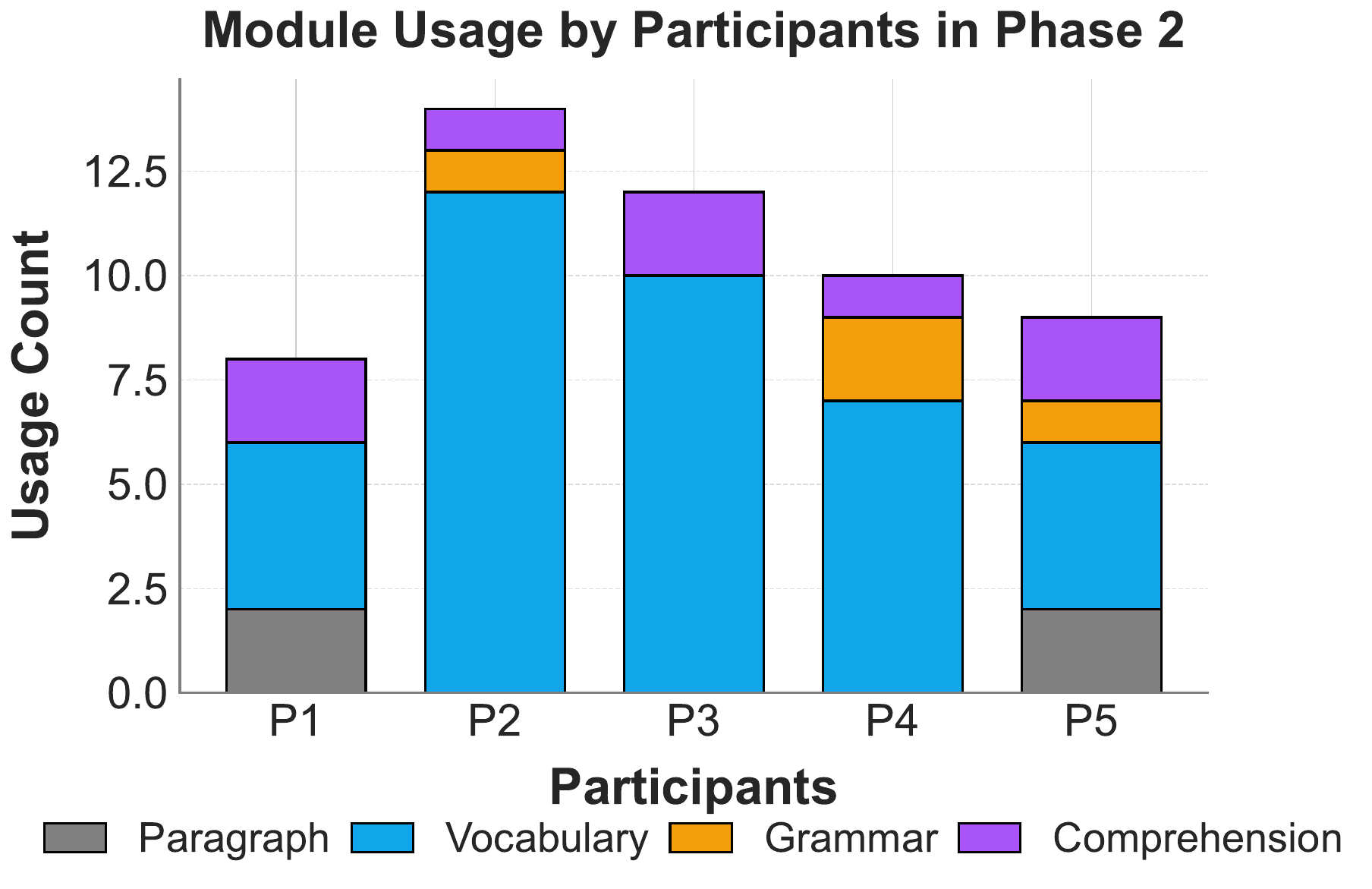}
    \caption{Usage of modules in \techname{} during the experiment by participants (P1-P5). We find that participants who used \techname{} used the tool mostly for searching for vocabulary. }
    \Description{This figure shows module usage statistics from an experimental study involving participants (P1-P5). The graph indicates that participants primarily utilized Reading.help for vocabulary searches, while also tracking the use of other modules during the session.}
    \label{fig:overview-chart}
\end{figure}

\begin{table*}[ht]
\centering
\small
\caption{Precision, recall, and F1 scores of keywords recommended by interpretable CEFR prediction models, with respect to the words and sentences that the participants in our study found it difficult. The analysis of the table is described in \S~\ref{subsec:results-readinghelp}.}
\resizebox{\linewidth}{!}{%
\begin{tabular}{l@{\hspace{8pt}}cccccc@{\hspace{8pt}}cccccc@{\hspace{8pt}}cccccc}
\toprule
 & \multicolumn{6}{c}{Recall (\%)} & \multicolumn{6}{c}{Precision (\%)} & \multicolumn{6}{c}{F1 Score (\%)} \\
\cmidrule(lr){2-7} \cmidrule(lr){8-13} \cmidrule(lr){14-19}
             & R1 & R2 & R3 & R4 & R5 & Avg. 
             & R1 & R2 & R3 & R4 & R5 & Avg. 
             & R1 & R2 & R3 & R4 & R5 & Avg. \\
\midrule
\makecell[r]{\textbf{B2 $\geq$}}
 & \textbf{75.00} & \textbf{81.25} & \textbf{87.50} & \textbf{94.44} & \textbf{100} & \textbf{87.64}
 & 18.37 & 19.40 & 19.44 & 20.24 & \textbf{14.89} & 18.47
 & 29.51 & \textbf{31.33} & 31.82 & 33.33 & \textbf{25.93} & 30.38 \\
\makecell[r]{\textbf{C1 $\geq$}}
 & 38.46 & 27.27 & 50.00 & 81.25 & 42.86 & 47.97
 & \textbf{35.71} & \textbf{31.58} & \textbf{45.00} & \textbf{65.00} & 11.54 & \textbf{37.77}
 & \textbf{37.04} & 29.27 & \textbf{47.37} & \textbf{72.22} & 18.18 & \textbf{40.82} \\
\bottomrule
\end{tabular}
}
\Description{This table displays the performance metrics—precision, recall, and F1 scores—of keywords recommended by CEFR models. The evaluation compares the system’s recommendations with the participants’ highlighted difficult sentences, considering a match when one highlighted portion is a subset of the other.}
\label{table:accuracy-llm}
\end{table*}

\smallskip
\noindent\textbf{Usage patterns. }Overall, 5 participants highlighted a total of 101 challenging parts of the text across five texts. 
\autoref{fig:overview-chart} shows an overview of which modules participants used in \techname{}. 
We find that participants use vocabulary models the modules the most, followed by comprehension, and grammar modules.
Participants' comments indicate that the comprehension module aided their understanding of uncertain sentences. 
It explained not only key terms but also provided background context for the entire sentence. 
The grammar component clarified ambiguous interpretations, and two participants noted that the paragraph summary helped them read and follow the text more swiftly.

\smallskip

\noindent\textbf{Assessment of CEFR model recommendations. } We evaluate interpretable CEFR prediction models on their ability to recommend text that readers might not know. 
\autoref{table:accuracy-llm} reports recall, precision, and F1 across five readings (R1–R5) under two difficulty thresholds, \textbf{$\geq$B2} and \textbf{$\geq$C1}. Because participants could freely highlight text without specifying whether they were unsure about word meaning or overall context, we employed token-based Jaccard similarity to measure semantic overlap between each recommendation and the union of user highlights. For each AI recommendation, we tokenized text by word boundaries and computed the Jaccard coefficient (intersection over union of word sets) against all user-highlighted content. A match was determined when the similarity exceeded $\theta = 0.3$.

At the \textbf{B2 $\geq$} threshold, the model achieves very high recall (75.00–100\%, Avg.\ 87.64\%) with low precision (14.89–20.24\%, Avg.\ 18.47\%), yielding an average F1 of 30.38\%. At the stricter \textbf{C1 $\geq$} threshold, precision improves (11.54–65.00\%, Avg.\ 37.77\%) while recall is lower and more variable (27.27–81.25\%, Avg.\ 47.97\%), resulting in a higher average F1 of 40.82\% with a peak of 72.22\% on R4. These results indicate a clear balance between coverage and specificity. The B2 $\geq$ threshold favors coverage by recovering most difficult items that participants highlighted. The C1 $\geq$ threshold favors specificity by reducing false positives and improving overall balance as reflected in the higher F1.

\begin{table*}[htbp]
\centering
\small
\caption{Confusion pattern types and examples of \techname{} recommendations that matched (or unmatched) actual EFL learner confusions.}
\resizebox{0.9\linewidth}{!}{%
\begin{tabular}{@{}lllll@{}}
\toprule
\textbf{Confusion Type} & \textbf{Explanation} & \textbf{Type} & \textbf{Detected} & \textbf{Missed} \\
\midrule
\multirow{4.5}{*}{C1: Contextual terms} &
\multirow{4.5}{*}{\makecell[l]{Uncommon, technical, or\\domain-specific terms}} &
R1 & \textit{Grunnings} / \textit{tawny} & \textit{unDursleyish} \\ \cmidrule(lr){3-5}
& & R3 & \textit{deterrence} / \textit{spectacle} / \textit{calculus} / \textit{arsenal} / \textit{flashpoint} & \textit{nukes} \\ \cmidrule(lr){3-5}
& & R5 & \textit{motifs} / \textit{detection} & \textit{-} \\ \cmidrule(lr){1-5}
\multirow{4.2}{*}{C2: Idioms \& Figurative} &
\multirow{4.2}{*}{\makecell[l]{Metaphorical or \\non-literal expressions}} &
R1 & \textit{didn't hold with such nonsense} & — \\ \cmidrule(lr){3-5}
& & \multirow{3}{*}{R3} & \textit{toyed with} / & \multirow{3}{*}{\textit{talked down}}\\
& &  & \textit{The idea has gone from fringe to mainstream} / & \\
& &  & \textit{may be an explosive political spectacle} & \\ \cmidrule(lr){1-5}
\multirow{2.4}{*}{C3: Polysemous words} &
\multirow{2.4}{*}{\makecell[l]{Words with multiple\\context-sensitive meanings}} &
R1 & \textit{dull} & \textit{bear} \\ \cmidrule(lr){3-5}
& & R3 & \textit{fringe} / \textit{reassure} & — \\
\bottomrule
\end{tabular}
}
\Description{This table categorizes different confusion pattern types observed among EFL learners and provides examples of Reading.help recommendations. The table indicates instances where the recommendations either matched or missed the actual learner confusions.}
\label{tab:ai_confusion_examples}
\end{table*}

\smallskip
\noindent\textbf{Qualitative assessment of CEFR model recommendations. } To better understand the strengths and limitations of our AI-based recommendation system, we conducted a pattern analysis using data from three reading passages: Reading 1 (R1), Reading 3 (R3), and Reading 5 (R5). 
We compared the language features that EFL readers highlighted as confusing with those that the system automatically recommended. 

\autoref{tab:ai_confusion_examples} shows the examples of keywords/phrases that the CEFR models predicted to be what the EFL readers may not know. 
We identify three patterns about CEFR models based on the analysis of keywords that are detected and missed. 
This comparative analysis revealed three dominant types of learner confusion: rare or invented terms (C1), idiomatic and figurative expressions (C2), and polysemous common words (C3).

First, under CEFR-based recommendation (>B2), the system retrieves many context-related terms but still misses a few, depending on text type. It correctly surfaced items such as \textit{Grunnings} and \textit{tawny} (R1), and from R3: \textit{deterrence}, \textit{spectacle}, \textit{calculus}, \textit{arsenal}, and \textit{flashpoint}; R5 yielded \textit{motifs} and \textit{detection}. In contrast, nonce or short forms like \textit{unDursleyish} (R1) and \textit{nukes} (R3) were “missed.” Given that \textit{unDursleyish} lacks a stable lexical entry, which makes CEFR estimation uncertain, we heuristically assign it B2; both terms therefore fall at or below our >B2 gate and are not recommended by design.

Second, for idioms and figurative language, the system captured expressions such as \textit{didn’t hold with such nonsense} (R1), \textit{toyed with}, \textit{The idea has gone from fringe to mainstream}, and \textit{may be an explosive political spectacle} (R3). The phrase \textit{talked down} (R3) was not captured, but it is B1 and thus intentionally filtered out under the current threshold.

Third, the system struggled with polysemous, high-frequency words whose meanings shift in context. It identified \textit{dull} (R1) and, in R3, \textit{fringe} and \textit{reassure}, but it missed \textit{bear} (B2) in “They didn’t think they could bear it if anyone found out about the Potters.” Because B2 items sit at or below our recommendation cutoff (>B2), this miss does not affect recommendations; however, it highlights that lower-CEFR words can still be confusing when used in marked senses or collocations.

Overall, CEFR gating usefully reduces oversuggestion and focuses attention on clearly challenging (C1–C2) items, especially domain-specific and idiomatic expressions. A remaining limitation is sensitivity to contextually tricky but lower-CEFR words; augmenting CEFR-based filtering with sense-aware cues could better flag these cases.

\smallskip
\noindent \textbf{Assessment of LLM Validity. } To identify \techname{}'s response quality, three English experts were recruited to evaluate a total of 300 examples. 
Responses deemed error-free were labeled ``Valid,'' and those containing errors ``Invalid,''; experts recorded the error types focusing on hallucinations and incorrect explanations and provided detailed justifications for each judgment. After this human-led evaluation, a self-validation process was conducted using an LLM, and we compared the results.

As shown in \autoref{tab:validation_counts}, the human experts' assessments revealed an overall valid rate of 88\% for vocabulary, 93\% for comprehension, and 90\% for grammar. 
Vocabulary emerged as the least accurate component, with 12\% of the responses flagged for errors. 
Incorrect translation of vocabulary accounted for 8 cases. Upon rechecking these eight cases, six involved semantically equivalent Korean paraphrases, whereas two were judged completely incorrect. Hence, while there exist some minor errors, we want to highlight that true semantic errors were rare. Comprehension errors (7\% invalid) mainly involved reasoning that extended beyond the scope of the given sentences, while grammar errors (10\% invalid) frequently related to incorrect parts of speech.

\smallskip
\noindent \textbf{LLM's self validation vs. human validation. }The LLM’s self-validation rated the responses as 93\% Valid for vocabulary, 97\% Valid for comprehension, and 96\% valid for Grammar. 
According to the LLM, vocabulary mistakes (7\% invalid) were often due to fabricated or irrelevant contexts and mistranslations, while comprehension issues (3\% invalid) stemmed from fabricated details beyond the original text. 
Grammar errors (4\% invalid) centered on incorrect verb tenses. 
Except for omissions of Korean translations in Vocabulary, the LLM’s error diagnoses largely aligned with those of the human experts.

In addition to validity rates, \autoref{tab:validation_accuracy} presents accuracy metrics for vocabulary, comprehension, and grammar tasks by self-validation. 
All metrics (Precision, Recall, F1 Score) were calculated using `valid' as the positive class, with human expert evaluations serving as the ground truth.

Comprehension shows the highest agreement with human judgments. Vocabulary is similar to grammar and slightly higher. Overall, these findings indicate that LLMs and humans exhibit similar validation behavior.

\vspace{-6pt}

\begin{table}[tbp]
\centering
\caption{Accuracy metrics for Vocabulary, Comprehension, and Grammar tasks by Self-Validation. All metrics (precision, recall, F1) were calculated using `valid' as the positive class, with human expert evaluations serving as the ground truth.}
\resizebox{0.9\columnwidth}{!}{%
\begin{tabular}{l cccc}
\toprule
 & Precision & Recall & F1 Score & Accuracy \\
\midrule
Vocabulary & 91.4 & 96.59 & 93.92 & 89 \\
Comprehension & 93.81 & 97.85 & 95.79 & 92 \\
Grammar & 90.63 & 96.67 & 93.55 & 88 \\
\bottomrule
\end{tabular}
}
\Description{This table shows the accuracy metrics—precision, recall, and F1 scores—for Vocabulary, Comprehension, and Grammar tasks based on a self-validation process. In these evaluations, ‘valid’ responses are the positive class, and human expert evaluations serve as the ground truth.}
\label{tab:validation_accuracy}
\end{table}

\begin{table}[tbp]
\footnotesize
\centering
\caption{Comparison of valid and invalid response rates (in \%) across Vocabulary, Comprehension, and Grammar, as evaluated by both human experts and the LLM.}
\resizebox{1.0\columnwidth}{!}{%
\begin{tabular}{l*{6}{c}}
\toprule
 & \multicolumn{2}{c}{Vocab} & \multicolumn{2}{c}{Comprehension} & \multicolumn{2}{c}{Grammar} \\
\cmidrule(lr){2-3} \cmidrule(lr){4-5} \cmidrule(lr){6-7}
 & \makecell{Valid} & \makecell{Invalid} & \makecell{Valid} & \makecell{Invalid} & \makecell{Valid} & \makecell{Invalid} \\
\midrule
\makecell{Human Expert} & 88\% & 12\% & 93\% & 7\% & 90\% & 10\% \\
LLM (GPT-4o) & 93\% & 7\% & 97\% & 3\% & 96\% & 4\% \\
\bottomrule
\end{tabular}
}
\Description{This table compares the response rates—expressed as percentages of valid versus invalid responses—for Vocabulary, Comprehension, and Grammar tasks. The table includes assessments from both human experts and the LLM, highlighting differences in evaluation outcomes.}
\label{tab:validation_counts}
\end{table}

\subsection{Expert feedback}
After the experiment, we asked two domain experts, one high-school English teacher, and another expert who is a certified English teacher, but working at a startup on online English education company.
We first asked them about the effectiveness of the tool, and then about the ideal role of a reading assistant for self-studying purposes for EFL readers. 
Below we describe their opinions. 

\smallskip
\noindent \textbf{Effectiveness of \techname{}.} To begin with, we identify four benefits of using our tool.
A primary advantage is its capacity to mitigate text complexity issues and enhance comprehension. 
Consequently, it helps the user to grasp not only the details but also the overall structure of the text.
One of the experts had never encountered such experience while reading, and expressed the willingness to test reading the entire fiction using \techname{}.
The second benefit is its ability to provide recommendations regarding topics that the user may find challenging. 
It is acknowledged that from a learner’s perspective, it is not always easy to know what one does not know, and such recommendations help address this issue.
Third, a significant advantage is the freedom to make unlimited requests at any time without pressure. 
Since \techname{} is a machine and not a human, not only does it respond to requests synchronously, but the user is also free to ask as many questions as desired without interference.
Fourth, it provides information tailored to the user's English proficiency level. 
Both experts argue that they expect such a tool to offer adaptive feedback, which is a key attribute of an effective human tutor.

\smallskip
\noindent \textbf{Improvement areas needed on \techname{}. }
The experts also identified three issues that must be addressed to improve service for EFL readers. 
First, they noted that the tool is effective only for individuals who are highly motivated to self-study English, rather than for those lacking sufficient motivation. 
To attract less motivated users, the tool should be made more accessible. 
For example, by offering additional recommendations and visual highlights. Second, the tool is not fully tailored to individual users. 
It could provide a more detailed analysis of each learner’s condition, offer an assessment of their English proficiency, and deliver explanations that are better customized to their needs. 
Although the current interface incorporates some of these features, both experts did not consider them sufficient. 
Finally, because the content is generated using LLMs, teachers acknowledge that while it can produce interesting material, verifying its accuracy remains challenging. 
In this regard, teachers expressed the need for a tool that they can use without having to worry about constant validation.

\section{Discussions}
\label{sec:07-discussion}

We describe the discussions based on the observation of the results. 
To begin with, we interpret the results (\S~\ref{subsec:07-1-implication}). 
Then, we discuss, in detail, the lessons learned from the study (\S~\ref{subsec:adaptiveness}, \S~\ref{subsec:engagement}, and \S~\ref{subsec:augmenting}.)

\subsection{Implication of Results} %
\label{subsec:07-1-implication}

Below, we present the implications from two parts: (1) how the tool is used, and the CEFR model's recommendation and validation performances of LLMs.

From the results, we find out that while reading a 400-word script, they most of the time look for vocabulary, on 5, to 6 times on average, and once or twice in comprehension.
We also find that occasionally, they used grammar and looked for paragraph summaries. 
This represents that in general, for EFL readers to fluently understand the text as well as the context (see \S~\ref{subsec:ESL-difficulties}), it requires a lot of effort. 
This showcases the utility of providing proactive guidance for EFL readers.

While both experts and participants acknowledged the effectiveness of the LLMs, there is still some issues on the validation of the contents. 
For this, we assess our system from two ways (1) how well their recommendations adhere to those of humans, and (2) how factual the responses are. 
Our recommendation provides 
From our results, we observed approximately 37.77\% precision when recommending a few items (see \autoref{table:accuracy-llm}). 
However, by increasing the number of recommendations, the tool can capture a larger portion of the information that EFL readers may not know. 
This is supported by the high recall rate—up to 87.64\%—observed in the B2 $\geq$ configuration in \autoref{table:accuracy-llm}. 
In summary, a recommendation strategy similar to the B2 $\geq$ configuration can facilitate the reading process by suggesting many keywords that the user may not know.

\subsection{Designing Adaptiveness for EFL Learning}
\label{subsec:adaptiveness}

A new level of adaptiveness, as mentioned by the experts, can deliver more precise and efficient support for EFL readers.   
A developed version of \techname{} may enlighten the users with faster, proactive guidance on unfamiliar word meanings, grammar, and comprehension. 
Both the language of guidance (e.g., level of explanation, text organization) and the type of feedback can be personalized based on analysis of a learner's accumulated skills. 
Such enhancements can accelerate reading and build confidence.

However, we should also be aware of two potential dangers: extra mental load and over-reliance. 
Excessive explanation can overwhelm the users, and replace the productive effort that builds skill. 
This is suggested by how participants took advantage of detailed grammatical information in \techname{} (see \S~\ref{subsec:results-readinghelp}---participants were overwhelmed with the amount of text, and later on did not again refer to the part). 
We therefore recommend brief and lightweight by default.
If providing details is necessary, then first hide detailed information, then show details only if the user chooses to.

Over-reliance is the second concern. 
The ultimate goal of reading tools like \techname{} is in increasing the user's English skills, so that reliance on the tool decreases as proficiency grows.
Consistent with the ``no pain, no gain'' principle in learning~\cite{debruin2023worth}, guidance should not be provided in a way that requires least amount of effort from the users. 
Instead, systems should adopt policies that deliberately preserve productive struggle to some extent. 
For example, withholding hints for terms that have been encountered repeatedly or gradually fading support. Success for tools like \techname{} should be measured not only by faster reading but also by growing independence over time.

\subsection{Engagement Strategies for EFL Readers}
\label{subsec:engagement}

Domain experts emphasized that higher engagement can sustain attention and improve learning outcomes for EFL readers. 
One way to increase engagement is to actively attract the user’s interest. 
In data visualization, for example, researchers have studied visual embellishment~\cite{bateman10chartjunk,parsons20chartjunk}—adding pictures or icons to charts to make them stand out. 
These additions can draw attention and encourage exploration, but prior work shows they can also reduce how clearly the data is understood. 
This trade-off has led some to call overly decorative elements \textit{chartjunk}. 
The same risk applies to reading assistance tools: methods that focus mainly on catching the eye may not improve understanding or skill development.
We therefore treat interest‑driven cues as tools to be used sparingly and with clear task alignment. 
For example, tools themselves can support motivation in the form of timely, lightweight texts, or progress bars~\cite{jeong25cpg}. 
The key is to channel attention toward the learning objective rather than to compete for it.

An alternative is to raise engagement by removing distraction. 
Rather than competing for attention, we envision an immersive~\cite{lee2025subg,lee2024productivity,shin24situatedsuvey}, purpose‑built environment for deep reading, paired with a human‑like virtual English tutor. 
Within this controlled space, the tutor offers just-in-time, ad-hoc clarification, and examples tailored to the reader's current difficulty, without interruptions that could potentially disturb comprehension. 
The system helps by making the reader's focus remains on the text by minimizing opportunities for attention shifts, and at the same time providing visible and accessible support.
We identify various challenges along this strategy, such as how to calibrate guidance and pace interventions, and validation. 

To that end, we propose a deep assessment of these two strategies-interest attraction versus distraction‑free immersion, for future development of foreign language education using computing devices.

\subsection{Augmenting, Not Replacing: The Role of LLMs in English Reading Education} 
\label{subsec:augmenting}

With the power of generative models, we can envision an English reading assistant that has absorbed a vast range of texts, can detect the specific difficulties a learner faces, and can deliver content at the learner’s level of proficiency. 
Such a tool could also generate and grade questions to check comprehension, raising the question: What would the ideal English reading assistant look like, and could it fully replace human teachers?

Despite these possibilities, there remain essential roles that only humans can fulfill in teaching English reading. 
Teachers are accountable for learner progress in a way that no generative models can be. 
They provide motivation, encouragement, and adaptive care—especially for beginners or those with low self-efficacy—when learners tire or lose momentum. 
Moreover, language learning is not only about mastering reading skills but also about building collaboration and social interaction, dimensions that require authentic human presence.

In this light, we envision an LLM-powered English reading assistant that augments rather than replaces teachers. 
For any text a learner brings, the assistant could offer adaptive support and remain available anytime and anywhere, even when instructors are not present. 
At the same time, when teachers are available, the tool could enhance their pedagogical decision-making and amplify their ability to guide students. 
Looking ahead, it is crucial to design such tools to advance English learning while preserving and strengthening the agency of both teachers and learners—an important direction for future work.

\section{Conclusion}
\label{sec:07-conclusion}

In this work, we developed a tool, called \techname{} to help EFL readers' reading processes.
In doing so, we utilized LLMs and interpretable neural networks. 
From our initial design, \techname{}-Lite, we conducted a pilot study, and based on it we developed a tool that effectively helps EFL readers, the \techname{}. 
From our experiment, we also found that CEFR models are capable of effectively recommending parts that EFL readers do not know. 
Moreover, LLMs possess abilities, though limited, to verify if the contents are incorrect.

\begin{acks}
We thank the anonymous reviewers for their precious feedback. 
We also thank Jinyoung Lee and Minhye Park for their precious feedback regarding the work. 
\end{acks}

\bibliographystyle{ACM-Reference-Format}
\bibliography{studies-for-babel}


\begin{thebibliography}{76}


\ifx \showCODEN    \undefined \def \showCODEN     #1{\unskip}     \fi
\ifx \showDOI      \undefined \def \showDOI       #1{#1}\fi
\ifx \showISBNx    \undefined \def \showISBNx     #1{\unskip}     \fi
\ifx \showISBNxiii \undefined \def \showISBNxiii  #1{\unskip}     \fi
\ifx \showISSN     \undefined \def \showISSN      #1{\unskip}     \fi
\ifx \showLCCN     \undefined \def \showLCCN      #1{\unskip}     \fi
\ifx \shownote     \undefined \def \shownote      #1{#1}          \fi
\ifx \showarticletitle \undefined \def \showarticletitle #1{#1}   \fi
\ifx \showURL      \undefined \def \showURL       {\relax}        \fi
\providecommand\bibfield[2]{#2}
\providecommand\bibinfo[2]{#2}
\providecommand\natexlab[1]{#1}
\providecommand\showeprint[2][]{arXiv:#2}

\bibitem[Afiliani et~al\mbox{.}(2024)]%
        {afiliani2024use}
\bibfield{author}{\bibinfo{person}{Afiliani Afiliani}, \bibinfo{person}{Inggrit~O Tanasale}, {and} \bibinfo{person}{Helena~M Rijoly}.} \bibinfo{year}{2024}\natexlab{}.
\newblock \showarticletitle{The Use of Google Translate in the Translation Class at English Education Study Program Pattimura University}.
\newblock \bibinfo{journal}{\emph{Journal of English Language Teaching and Linguistics}} \bibinfo{volume}{9}, \bibinfo{number}{1} (\bibinfo{year}{2024}), \bibinfo{pages}{95--109}.
\newblock


\bibitem[Aleksandrova and Pouliot(2023)]%
        {aleksandrova2023cefr}
\bibfield{author}{\bibinfo{person}{Desislava Aleksandrova} {and} \bibinfo{person}{Vincent Pouliot}.} \bibinfo{year}{2023}\natexlab{}.
\newblock \showarticletitle{Cefr-based contextual lexical complexity classifier in english and french}. In \bibinfo{booktitle}{\emph{Proceedings of the 18th Workshop on Innovative Use of NLP for Building Educational Applications (BEA 2023)}}. \bibinfo{pages}{518--527}.
\newblock


\bibitem[Arakawa et~al\mbox{.}(2022)]%
        {arakawa2022vocabencounter}
\bibfield{author}{\bibinfo{person}{Riku Arakawa}, \bibinfo{person}{Hiromu Yakura}, {and} \bibinfo{person}{Sosuke Kobayashi}.} \bibinfo{year}{2022}\natexlab{}.
\newblock \showarticletitle{VocabEncounter: NMT-powered Vocabulary Learning by Presenting Computer-Generated Usages of Foreign Words into Users’ Daily Lives}. In \bibinfo{booktitle}{\emph{Proceedings of the ACM Conference on Human Factors in Computing Systems (CHI)}}. \bibinfo{publisher}{Association for Computing Machinery}, \bibinfo{address}{New York, NY, USA}, Article \bibinfo{articleno}{6}, \bibinfo{numpages}{21}~pages.
\newblock
\urldef\tempurl%
\url{https://doi.org/10.1145/3491102.3501839}
\showDOI{\tempurl}


\bibitem[Arase et~al\mbox{.}(2022)]%
        {arase2022cefr}
\bibfield{author}{\bibinfo{person}{Yuki Arase}, \bibinfo{person}{Satoru Uchida}, {and} \bibinfo{person}{Tomoyuki Kajiwara}.} \bibinfo{year}{2022}\natexlab{}.
\newblock \showarticletitle{CEFR-based sentence difficulty annotation and assessment}.
\newblock \bibinfo{journal}{\emph{arXiv preprint arXiv:2210.11766}} (\bibinfo{year}{2022}).
\newblock


\bibitem[Ba et~al\mbox{.}(2016)]%
        {ba2016layernormalization}
\bibfield{author}{\bibinfo{person}{Jimmy~Lei Ba}, \bibinfo{person}{Jamie~Ryan Kiros}, {and} \bibinfo{person}{Geoffrey~E. Hinton}.} \bibinfo{year}{2016}\natexlab{}.
\newblock \bibinfo{title}{Layer Normalization}.
\newblock
\newblock
\showeprint[arxiv]{1607.06450}~[stat.ML]
\urldef\tempurl%
\url{https://arxiv.org/abs/1607.06450}
\showURL{%
\tempurl}


\bibitem[Badam et~al\mbox{.}(2019)]%
        {badam2019elasticdocuments}
\bibfield{author}{\bibinfo{person}{Sriram~Karthik Badam}, \bibinfo{person}{Zhicheng Liu}, {and} \bibinfo{person}{Niklas Elmqvist}.} \bibinfo{year}{2019}\natexlab{}.
\newblock \showarticletitle{Elastic Documents: Coupling Text and Tables through Contextual Visualizations for Enhanced Document Reading}.
\newblock \bibinfo{journal}{\emph{IEEE Transactions on Visualization and Computer Graphics}} \bibinfo{volume}{25}, \bibinfo{number}{1} (\bibinfo{date}{Jan} \bibinfo{year}{2019}), \bibinfo{pages}{661--671}.
\newblock
\urldef\tempurl%
\url{https://doi.org/10.1109/TVCG.2018.2865119}
\showDOI{\tempurl}


\bibitem[Bateman et~al\mbox{.}(2010)]%
        {bateman10chartjunk}
\bibfield{author}{\bibinfo{person}{Scott Bateman}, \bibinfo{person}{Regan~L. Mandryk}, \bibinfo{person}{Carl Gutwin}, \bibinfo{person}{Aaron Genest}, \bibinfo{person}{David McDine}, {and} \bibinfo{person}{Christopher Brooks}.} \bibinfo{year}{2010}\natexlab{}.
\newblock \showarticletitle{Useful junk? the effects of visual embellishment on comprehension and memorability of charts}. In \bibinfo{booktitle}{\emph{Proceedings of the ACM Conference on Human Factors in Computing Systems}}. \bibinfo{publisher}{Association for Computing Machinery}, \bibinfo{address}{New York, NY, USA}, \bibinfo{pages}{2573–2582}.
\newblock
\showISBNx{9781605589299}
\urldef\tempurl%
\url{https://doi.org/10.1145/1753326.1753716}
\showDOI{\tempurl}


\bibitem[Carr(1981)]%
        {carr1981readingability}
\bibfield{author}{\bibinfo{person}{Thomas~H. Carr}.} \bibinfo{year}{1981}\natexlab{}.
\newblock \showarticletitle{Building theories of reading ability: On the relation between individual differences in cognitive skills and reading comprehension}.
\newblock \bibinfo{journal}{\emph{Cognition}} \bibinfo{volume}{9}, \bibinfo{number}{1} (\bibinfo{year}{1981}), \bibinfo{pages}{73--114}.
\newblock
\urldef\tempurl%
\url{https://doi.org/10.1016/0010-0277(81)90015-9}
\showDOI{\tempurl}


\bibitem[Chang et~al\mbox{.}(2023)]%
        {chang2023citesee}
\bibfield{author}{\bibinfo{person}{Joseph~Chee Chang}, \bibinfo{person}{Amy~X. Zhang}, \bibinfo{person}{Jonathan Bragg}, \bibinfo{person}{Andrew Head}, \bibinfo{person}{Kyle Lo}, \bibinfo{person}{Doug Downey}, {and} \bibinfo{person}{Daniel~S. Weld}.} \bibinfo{year}{2023}\natexlab{}.
\newblock \showarticletitle{CiteSee: Augmenting Citations in Scientific Papers with Persistent and Personalized Historical Context}. In \bibinfo{booktitle}{\emph{Proceedings of the ACM Conference on Human Factors in Computing Systems (CHI)}}. \bibinfo{publisher}{Association for Computing Machinery}, \bibinfo{address}{New York, NY, USA}, Article \bibinfo{articleno}{737}, \bibinfo{numpages}{15}~pages.
\newblock
\urldef\tempurl%
\url{https://doi.org/10.1145/3544548.3580847}
\showDOI{\tempurl}


\bibitem[Chen et~al\mbox{.}(2023)]%
        {chen2023marvista}
\bibfield{author}{\bibinfo{person}{Xiang~“Anthony” Chen}, \bibinfo{person}{Chien-Sheng Wu}, \bibinfo{person}{Lidiya Murakhovs’ka}, \bibinfo{person}{Philippe Laban}, \bibinfo{person}{Tong Niu}, \bibinfo{person}{Wenhao Liu}, {and} \bibinfo{person}{Caiming Xiong}.} \bibinfo{year}{2023}\natexlab{}.
\newblock \showarticletitle{Marvista: Exploring the Design of a Human-AI Collaborative News Reading Tool}.
\newblock \bibinfo{journal}{\emph{ACM Trans. Comput.-Hum. Interact.}} \bibinfo{volume}{30}, \bibinfo{number}{6}, Article \bibinfo{articleno}{92} (\bibinfo{date}{sep} \bibinfo{year}{2023}), \bibinfo{numpages}{27}~pages.
\newblock
\urldef\tempurl%
\url{https://doi.org/10.1145/3609331}
\showDOI{\tempurl}


\bibitem[Choi et~al\mbox{.}(2019)]%
        {choi19visblind}
\bibfield{author}{\bibinfo{person}{Jinho Choi}, \bibinfo{person}{Sanghun Jung}, \bibinfo{person}{Deok~Gun Park}, \bibinfo{person}{Jaegul Choo}, {and} \bibinfo{person}{Niklas Elmqvist}.} \bibinfo{year}{2019}\natexlab{}.
\newblock \showarticletitle{Visualizing for the Non-Visual: Enabling the Visually Impaired to Use Visualization}.
\newblock \bibinfo{journal}{\emph{Computer Graphics Forum}} \bibinfo{volume}{38}, \bibinfo{number}{3} (\bibinfo{year}{2019}), \bibinfo{pages}{249--260}.
\newblock
\urldef\tempurl%
\url{https://doi.org/10.1111/cgf.13686}
\showDOI{\tempurl}


\bibitem[Chundury et~al\mbox{.}(2024)]%
        {chundary2024tactualplot}
\bibfield{author}{\bibinfo{person}{Pramod Chundury}, \bibinfo{person}{Yasmin Reyazuddin}, \bibinfo{person}{J.~Bern Jordan}, \bibinfo{person}{Jonathan Lazar}, {and} \bibinfo{person}{Niklas Elmqvist}.} \bibinfo{year}{2024}\natexlab{}.
\newblock \showarticletitle{TactualPlot: Spatializing Data as Sound Using Sensory Substitution for Touchscreen Accessibility}.
\newblock \bibinfo{journal}{\emph{IEEE Transactions on Visualization and Computer Graphics}} \bibinfo{volume}{30}, \bibinfo{number}{1} (\bibinfo{date}{Jan} \bibinfo{year}{2024}), \bibinfo{pages}{836--846}.
\newblock
\urldef\tempurl%
\url{https://doi.org/10.1109/TVCG.2023.3326937}
\showDOI{\tempurl}


\bibitem[de~Bruin et~al\mbox{.}(2023)]%
        {debruin2023worth}
\bibfield{author}{\bibinfo{person}{Anique~BH de Bruin}, \bibinfo{person}{Felicitas Biwer}, \bibinfo{person}{Luotong Hui}, \bibinfo{person}{Erdem Onan}, \bibinfo{person}{Louise David}, {and} \bibinfo{person}{Wisnu Wiradhany}.} \bibinfo{year}{2023}\natexlab{}.
\newblock \showarticletitle{Worth the effort: The start and stick to desirable difficulties (S2D2) framework}.
\newblock \bibinfo{journal}{\emph{Educational Psychology Review}} \bibinfo{volume}{35}, \bibinfo{number}{2} (\bibinfo{year}{2023}), \bibinfo{pages}{41}.
\newblock


\bibitem[Diasti et~al\mbox{.}(2023)]%
        {diasti2023implementation}
\bibfield{author}{\bibinfo{person}{Krismalita~Sekar Diasti}, \bibinfo{person}{Cecilia~Titiek Murniati}, {and} \bibinfo{person}{Heny Hartono}.} \bibinfo{year}{2023}\natexlab{}.
\newblock \showarticletitle{The Implementation of KWL Strategy in EFL Students' Reading Comprehension.}
\newblock \bibinfo{journal}{\emph{Journal of English Teaching}} \bibinfo{volume}{9}, \bibinfo{number}{2} (\bibinfo{year}{2023}), \bibinfo{pages}{176--185}.
\newblock


\bibitem[D'Mello et~al\mbox{.}(2016)]%
        {dmello16mindwandering}
\bibfield{author}{\bibinfo{person}{Sidney D'Mello}, \bibinfo{person}{Kristopher Kopp}, \bibinfo{person}{Robert~Earl Bixler}, {and} \bibinfo{person}{Nigel Bosch}.} \bibinfo{year}{2016}\natexlab{}.
\newblock \showarticletitle{Attending to Attention: Detecting and Combating Mind Wandering during Computerized Reading}. In \bibinfo{booktitle}{\emph{Proceedings of the Extended Abstracts on Human Factors in Computing Systems (CHI)}}. \bibinfo{publisher}{Association for Computing Machinery}, \bibinfo{address}{New York, NY, USA}, \bibinfo{pages}{1661–1669}.
\newblock
\urldef\tempurl%
\url{https://doi.org/10.1145/2851581.2892329}
\showDOI{\tempurl}


\bibitem[Fok et~al\mbox{.}(2023)]%
        {fok2023scim}
\bibfield{author}{\bibinfo{person}{Raymond Fok}, \bibinfo{person}{Hita Kambhamettu}, \bibinfo{person}{Luca Soldaini}, \bibinfo{person}{Jonathan Bragg}, \bibinfo{person}{Kyle Lo}, \bibinfo{person}{Marti Hearst}, \bibinfo{person}{Andrew Head}, {and} \bibinfo{person}{Daniel~S Weld}.} \bibinfo{year}{2023}\natexlab{}.
\newblock \showarticletitle{Scim: Intelligent Skimming Support for Scientific Papers}. In \bibinfo{booktitle}{\emph{Proceedings of the International Conference on Intelligent User Interfaces (IUI)}}. \bibinfo{publisher}{Association for Computing Machinery}, \bibinfo{address}{New York, NY, USA}, \bibinfo{pages}{476–490}.
\newblock
\urldef\tempurl%
\url{https://doi.org/10.1145/3581641.3584034}
\showDOI{\tempurl}


\bibitem[Fujinuma and Hagiwara(2021)]%
        {fujinuma2021semi}
\bibfield{author}{\bibinfo{person}{Yoshinari Fujinuma} {and} \bibinfo{person}{Masato Hagiwara}.} \bibinfo{year}{2021}\natexlab{}.
\newblock \showarticletitle{Semi-supervised joint estimation of word and document readability}.
\newblock \bibinfo{journal}{\emph{arXiv preprint arXiv:2104.13103}} (\bibinfo{year}{2021}).
\newblock


\bibitem[Gaba et~al\mbox{.}(2023)]%
        {gaba2023langaugevisual}
\bibfield{author}{\bibinfo{person}{Aimen Gaba}, \bibinfo{person}{Vidya Setlur}, \bibinfo{person}{Arjun Srinivasan}, \bibinfo{person}{Jane Hoffswell}, {and} \bibinfo{person}{Cindy Xiong}.} \bibinfo{year}{2023}\natexlab{}.
\newblock \showarticletitle{Comparison Conundrum and the Chamber of Visualizations: An Exploration of How Language Influences Visual Design}.
\newblock \bibinfo{journal}{\emph{IEEE Transactions on Visualization and Computer Graphics}} \bibinfo{volume}{29}, \bibinfo{number}{1} (\bibinfo{date}{Jan} \bibinfo{year}{2023}), \bibinfo{pages}{1211--1221}.
\newblock
\showISSN{1941-0506}
\urldef\tempurl%
\url{https://doi.org/10.1109/TVCG.2022.3209456}
\showDOI{\tempurl}


\bibitem[Gero et~al\mbox{.}(2023)]%
        {gero2023socialdynamicswriting}
\bibfield{author}{\bibinfo{person}{Katy~Ilonka Gero}, \bibinfo{person}{Tao Long}, {and} \bibinfo{person}{Lydia~B Chilton}.} \bibinfo{year}{2023}\natexlab{}.
\newblock \showarticletitle{Social Dynamics of AI Support in Creative Writing}. In \bibinfo{booktitle}{\emph{Proceedings of the ACM Conference on Human Factors in Computing Systems (CHI)}}. \bibinfo{publisher}{Association for Computing Machinery}, \bibinfo{address}{New York, NY, USA}, Article \bibinfo{articleno}{245}, \bibinfo{numpages}{15}~pages.
\newblock
\urldef\tempurl%
\url{https://doi.org/10.1145/3544548.3580782}
\showDOI{\tempurl}


\bibitem[Gou et~al\mbox{.}(2024)]%
        {gou2024criticlargelanguagemodels}
\bibfield{author}{\bibinfo{person}{Zhibin Gou}, \bibinfo{person}{Zhihong Shao}, \bibinfo{person}{Yeyun Gong}, \bibinfo{person}{Yelong Shen}, \bibinfo{person}{Yujiu Yang}, \bibinfo{person}{Nan Duan}, {and} \bibinfo{person}{Weizhu Chen}.} \bibinfo{year}{2024}\natexlab{}.
\newblock \bibinfo{title}{CRITIC: Large Language Models Can Self-Correct with Tool-Interactive Critiquing}.
\newblock
\newblock
\showeprint[arxiv]{2305.11738}~[cs.CL]
\urldef\tempurl%
\url{https://arxiv.org/abs/2305.11738}
\showURL{%
\tempurl}


\bibitem[Graesser et~al\mbox{.}(1980)]%
        {graesser1920readingcomponents}
\bibfield{author}{\bibinfo{person}{Arthur~C. Graesser}, \bibinfo{person}{Nicholas~L. Hoffman}, {and} \bibinfo{person}{Leslie~F. Clark}.} \bibinfo{year}{1980}\natexlab{}.
\newblock \showarticletitle{Structural components of reading time}.
\newblock \bibinfo{journal}{\emph{Journal of Verbal Learning and Verbal Behavior}} \bibinfo{volume}{19}, \bibinfo{number}{2} (\bibinfo{year}{1980}), \bibinfo{pages}{135--151}.
\newblock
\urldef\tempurl%
\url{https://doi.org/10.1016/S0022-5371(80)90132-2}
\showDOI{\tempurl}


\bibitem[Heer et~al\mbox{.}(2023)]%
        {heer2023livingpapers}
\bibfield{author}{\bibinfo{person}{Jeffrey Heer}, \bibinfo{person}{Matthew Conlen}, \bibinfo{person}{Vishal Devireddy}, \bibinfo{person}{Tu Nguyen}, {and} \bibinfo{person}{Joshua Horowitz}.} \bibinfo{year}{2023}\natexlab{}.
\newblock \showarticletitle{Living Papers: A Language Toolkit for Augmented Scholarly Communication}. In \bibinfo{booktitle}{\emph{Proceedings of the ACM Symposium on User Interface Software and Technology (UIST)}}. \bibinfo{publisher}{Association for Computing Machinery}, \bibinfo{address}{New York, NY, USA}, Article \bibinfo{articleno}{42}, \bibinfo{numpages}{13}~pages.
\newblock
\urldef\tempurl%
\url{https://doi.org/10.1145/3586183.3606791}
\showDOI{\tempurl}


\bibitem[Hendrycks and Gimpel(2023)]%
        {hendrycks2023gaussianerrorlinearunits}
\bibfield{author}{\bibinfo{person}{Dan Hendrycks} {and} \bibinfo{person}{Kevin Gimpel}.} \bibinfo{year}{2023}\natexlab{}.
\newblock \bibinfo{title}{Gaussian Error Linear Units (GELUs)}.
\newblock
\newblock
\showeprint[arxiv]{1606.08415}~[cs.LG]
\urldef\tempurl%
\url{https://arxiv.org/abs/1606.08415}
\showURL{%
\tempurl}


\bibitem[Hinkel(2003)]%
        {hinkel2003teaching}
\bibfield{author}{\bibinfo{person}{Eli Hinkel}.} \bibinfo{year}{2003}\natexlab{}.
\newblock \bibinfo{booktitle}{\emph{Teaching academic ESL writing: Practical techniques in vocabulary and grammar}}.
\newblock \bibinfo{publisher}{Routledge}.
\newblock


\bibitem[Hirsch(2003)]%
        {hirsch2003reading}
\bibfield{author}{\bibinfo{person}{Eric~Donald Hirsch}.} \bibinfo{year}{2003}\natexlab{}.
\newblock \showarticletitle{Reading comprehension requires knowledge of words and the world}.
\newblock \bibinfo{journal}{\emph{American educator}} \bibinfo{volume}{27}, \bibinfo{number}{1} (\bibinfo{year}{2003}), \bibinfo{pages}{10--13}.
\newblock


\bibitem[Hong-Nam and Page(2014)]%
        {hong14investigating}
\bibfield{author}{\bibinfo{person}{Kay Hong-Nam} {and} \bibinfo{person}{Larkin Page}.} \bibinfo{year}{2014}\natexlab{}.
\newblock \showarticletitle{Investigating metacognitive awareness and reading strategy use of EFL Korean university students}.
\newblock \bibinfo{journal}{\emph{Reading Psychology}} \bibinfo{volume}{35}, \bibinfo{number}{3} (\bibinfo{year}{2014}), \bibinfo{pages}{195--220}.
\newblock


\bibitem[Hoque et~al\mbox{.}(2023)]%
        {hoque2023susurrus}
\bibfield{author}{\bibinfo{person}{Md~Naimul Hoque}, \bibinfo{person}{Md Ehtesham-Ul-Haque}, \bibinfo{person}{Niklas Elmqvist}, {and} \bibinfo{person}{Syed~Masum Billah}.} \bibinfo{year}{2023}\natexlab{}.
\newblock \showarticletitle{Accessible Data Representation with Natural Sound}. In \bibinfo{booktitle}{\emph{Proceedings of the ACM CHI Conference on Human Factors in Computing Systems (CHI)}}. \bibinfo{publisher}{Association for Computing Machinery}, \bibinfo{address}{New York, NY, USA}, Article \bibinfo{articleno}{826}, \bibinfo{numpages}{19}~pages.
\newblock
\urldef\tempurl%
\url{https://doi.org/10.1145/3544548.3581087}
\showDOI{\tempurl}


\bibitem[Hoque et~al\mbox{.}(2024a)]%
        {hoque2024hallmark}
\bibfield{author}{\bibinfo{person}{Md~Naimul Hoque}, \bibinfo{person}{Tasfia Mashiat}, \bibinfo{person}{Bhavya Ghai}, \bibinfo{person}{Cecilia~D. Shelton}, \bibinfo{person}{Fanny Chevalier}, \bibinfo{person}{Kari Kraus}, {and} \bibinfo{person}{Niklas Elmqvist}.} \bibinfo{year}{2024}\natexlab{a}.
\newblock \showarticletitle{The HaLLMark Effect: Supporting Provenance and Transparent Use of Large Language Models in Writing with Interactive Visualization}. In \bibinfo{booktitle}{\emph{Proceedings of the ACM Conference on Human Factors in Computing Systems (CHI)}}. \bibinfo{publisher}{Association for Computing Machinery}, \bibinfo{address}{New York, NY, USA}, Article \bibinfo{articleno}{1045}, \bibinfo{numpages}{15}~pages.
\newblock
\urldef\tempurl%
\url{https://doi.org/10.1145/3613904.3641895}
\showDOI{\tempurl}


\bibitem[Hoque et~al\mbox{.}(2024b)]%
        {hoque2024hcaitools}
\bibfield{author}{\bibinfo{person}{Md~Naimul Hoque}, \bibinfo{person}{Sungbok Shin}, {and} \bibinfo{person}{Niklas Elmqvist}.} \bibinfo{year}{2024}\natexlab{b}.
\newblock \showarticletitle{Visualization for human-centered AI tools}.
\newblock \bibinfo{journal}{\emph{arXiv preprint arXiv:2404.02147}} (\bibinfo{year}{2024}).
\newblock


\bibitem[Hu et~al\mbox{.}(2022)]%
        {hu2022relative}
\bibfield{author}{\bibinfo{person}{Tsui-Chun Hu}, \bibinfo{person}{Yao-Ting Sung}, \bibinfo{person}{Hsing-Huang Liang}, \bibinfo{person}{Tsung-Jen Chang}, {and} \bibinfo{person}{Yeh-Tai Chou}.} \bibinfo{year}{2022}\natexlab{}.
\newblock \showarticletitle{Relative roles of grammar knowledge and vocabulary in the reading comprehension of EFL elementary-school learners: Direct, mediating, and form/meaning-distinct effects}.
\newblock \bibinfo{journal}{\emph{Frontiers in psychology}}  \bibinfo{volume}{13} (\bibinfo{year}{2022}), \bibinfo{pages}{827007}.
\newblock


\bibitem[Ito et~al\mbox{.}(2023)]%
        {ito2024airewrite}
\bibfield{author}{\bibinfo{person}{Takumi Ito}, \bibinfo{person}{Naomi Yamashita}, \bibinfo{person}{Tatsuki Kuribayashi}, \bibinfo{person}{Masatoshi Hidaka}, \bibinfo{person}{Jun Suzuki}, \bibinfo{person}{Ge Gao}, \bibinfo{person}{Jack Jamieson}, {and} \bibinfo{person}{Kentaro Inui}.} \bibinfo{year}{2023}\natexlab{}.
\newblock \showarticletitle{Use of an AI-powered Rewriting Support Software in Context with Other Tools: A Study of Non-Native English Speakers}. In \bibinfo{booktitle}{\emph{Proceedings of the ACM Symposium on User Interface Software and Technology (UIST)}}. \bibinfo{publisher}{Association for Computing Machinery}, \bibinfo{address}{New York, NY, USA}, Article \bibinfo{articleno}{45}, \bibinfo{numpages}{13}~pages.
\newblock
\urldef\tempurl%
\url{https://doi.org/10.1145/3586183.3606810}
\showDOI{\tempurl}


\bibitem[Iwai(2008)]%
        {iwai2008perceptions}
\bibfield{author}{\bibinfo{person}{Yuko Iwai}.} \bibinfo{year}{2008}\natexlab{}.
\newblock \showarticletitle{The perceptions of Japanese students toward academic English reading: implications for effective ESL reading strategies.}
\newblock \bibinfo{journal}{\emph{Multicultural Education}} \bibinfo{volume}{15}, \bibinfo{number}{4} (\bibinfo{year}{2008}), \bibinfo{pages}{45--50}.
\newblock


\bibitem[Jeong et~al\mbox{.}(2025)]%
        {jeong25cpg}
\bibfield{author}{\bibinfo{person}{Daeun Jeong}, \bibinfo{person}{Sungbok Shin}, {and} \bibinfo{person}{Jongwook Jeong}.} \bibinfo{year}{2025}\natexlab{}.
\newblock \showarticletitle{Conversation Progress Guide: UI System for Enhancing Self-Efficacy in Conversational AI}. In \bibinfo{booktitle}{\emph{Proceedings of the ACM Conference on Human Factors in Computing Systems}}. \bibinfo{publisher}{Association for Computing Machinery}, \bibinfo{address}{New York, NY, USA}, Article \bibinfo{articleno}{180}, \bibinfo{numpages}{11}~pages.
\newblock
\urldef\tempurl%
\url{https://doi.org/10.1145/3706598.3714222}
\showDOI{\tempurl}


\bibitem[Joshi and Vogel(2024)]%
        {nikhita2024constrainedhigh}
\bibfield{author}{\bibinfo{person}{Nikhita Joshi} {and} \bibinfo{person}{Daniel Vogel}.} \bibinfo{year}{2024}\natexlab{}.
\newblock \showarticletitle{Constrained Highlighting in a Document Reader can Improve Reading Comprehension}. In \bibinfo{booktitle}{\emph{Proceedings of the ACM Conference on Human Factors in Computing Systems (CHI)}}. \bibinfo{publisher}{Association for Computing Machinery}, \bibinfo{address}{New York, NY, USA}, Article \bibinfo{articleno}{893}, \bibinfo{numpages}{10}~pages.
\newblock
\urldef\tempurl%
\url{https://doi.org/10.1145/3613904.3642314}
\showDOI{\tempurl}


\bibitem[Kang et~al\mbox{.}(2022a)]%
        {hang2022threddy}
\bibfield{author}{\bibinfo{person}{Hyeonsu Kang}, \bibinfo{person}{Joseph~Chee Chang}, \bibinfo{person}{Yongsung Kim}, {and} \bibinfo{person}{Aniket Kittur}.} \bibinfo{year}{2022}\natexlab{a}.
\newblock \showarticletitle{Threddy: An Interactive System for Personalized Thread-based Exploration and Organization of Scientific Literature}. In \bibinfo{booktitle}{\emph{Proceedings of the ACM Symposium on User Interface Software and Technology (UIST)}}. \bibinfo{publisher}{Association for Computing Machinery}, \bibinfo{address}{New York, NY, USA}, Article \bibinfo{articleno}{94}, \bibinfo{numpages}{15}~pages.
\newblock
\urldef\tempurl%
\url{https://doi.org/10.1145/3526113.3545660}
\showDOI{\tempurl}


\bibitem[Kang et~al\mbox{.}(2022b)]%
        {kang2022fwyktwyr}
\bibfield{author}{\bibinfo{person}{Hyeonsu~B Kang}, \bibinfo{person}{Rafal Kocielnik}, \bibinfo{person}{Andrew Head}, \bibinfo{person}{Jiangjiang Yang}, \bibinfo{person}{Matt Latzke}, \bibinfo{person}{Aniket Kittur}, \bibinfo{person}{Daniel~S Weld}, \bibinfo{person}{Doug Downey}, {and} \bibinfo{person}{Jonathan Bragg}.} \bibinfo{year}{2022}\natexlab{b}.
\newblock \showarticletitle{From Who You Know to What You Read: Augmenting Scientific Recommendations with Implicit Social Networks}. In \bibinfo{booktitle}{\emph{Proceedings of the ACM Conference on Human Factors in Computing Systems (CHI)}}. \bibinfo{publisher}{Association for Computing Machinery}, \bibinfo{address}{New York, NY, USA}, Article \bibinfo{articleno}{302}, \bibinfo{numpages}{23}~pages.
\newblock
\urldef\tempurl%
\url{https://doi.org/10.1145/3491102.3517470}
\showDOI{\tempurl}


\bibitem[Kang et~al\mbox{.}(2023)]%
        {kang2023comlittee}
\bibfield{author}{\bibinfo{person}{Hyeonsu~B Kang}, \bibinfo{person}{Nouran Soliman}, \bibinfo{person}{Matt Latzke}, \bibinfo{person}{Joseph~Chee Chang}, {and} \bibinfo{person}{Jonathan Bragg}.} \bibinfo{year}{2023}\natexlab{}.
\newblock \showarticletitle{ComLittee: Literature Discovery with Personal Elected Author Committees}. In \bibinfo{booktitle}{\emph{Proceedings of the ACM Conference on Human Factors in Computing Systems (CHI)}}. \bibinfo{publisher}{Association for Computing Machinery}, \bibinfo{address}{New York, NY, USA}, Article \bibinfo{articleno}{738}, \bibinfo{numpages}{20}~pages.
\newblock
\urldef\tempurl%
\url{https://doi.org/10.1145/3544548.3581371}
\showDOI{\tempurl}


\bibitem[Karolus et~al\mbox{.}(2023)]%
        {karolus2023hardtoread}
\bibfield{author}{\bibinfo{person}{Jakob Karolus}, \bibinfo{person}{Sebastian~S. Feger}, \bibinfo{person}{Albrecht Schmidt}, {and} \bibinfo{person}{Pawe\l{}~W. Wo\'{z}niak}.} \bibinfo{year}{2023}\natexlab{}.
\newblock \showarticletitle{Your Text Is Hard to Read: Facilitating Readability Awareness to Support Writing Proficiency in Text Production}. In \bibinfo{booktitle}{\emph{Proceedings of the ACM Designing Interactive Systems Conference (DIS)}}. \bibinfo{publisher}{Association for Computing Machinery}, \bibinfo{address}{New York, NY, USA}, \bibinfo{pages}{147–160}.
\newblock
\urldef\tempurl%
\url{https://doi.org/10.1145/3563657.3596052}
\showDOI{\tempurl}


\bibitem[Kelious et~al\mbox{.}(2024)]%
        {kelious2024complex}
\bibfield{author}{\bibinfo{person}{Abdelhak Kelious}, \bibinfo{person}{Matthieu Constant}, {and} \bibinfo{person}{Christophe Coeur}.} \bibinfo{year}{2024}\natexlab{}.
\newblock \showarticletitle{Complex word identification: A comparative study between ChatGPT and a dedicated model for this task}. In \bibinfo{booktitle}{\emph{Proceedings of the 2024 Joint International Conference on Computational Linguistics, Language Resources and Evaluation (LREC-COLING 2024)}}. \bibinfo{pages}{3645--3653}.
\newblock


\bibitem[Kim et~al\mbox{.}(2018)]%
        {kim2018doctexttable}
\bibfield{author}{\bibinfo{person}{Dae~Hyun Kim}, \bibinfo{person}{Enamul Hoque}, \bibinfo{person}{Juho Kim}, {and} \bibinfo{person}{Maneesh Agrawala}.} \bibinfo{year}{2018}\natexlab{}.
\newblock \showarticletitle{Facilitating Document Reading by Linking Text and Tables}. In \bibinfo{booktitle}{\emph{Proceedings of the ACM Symposium on User Interface Software and Technology (UIST)}}. \bibinfo{publisher}{Association for Computing Machinery}, \bibinfo{address}{New York, NY, USA}, \bibinfo{pages}{423–434}.
\newblock
\showISBNx{9781450359481}
\urldef\tempurl%
\url{https://doi.org/10.1145/3242587.3242617}
\showDOI{\tempurl}


\bibitem[Kim et~al\mbox{.}(2023)]%
        {kim2023treemapexplain}
\bibfield{author}{\bibinfo{person}{Gyeongri Kim}, \bibinfo{person}{Jiho Kim}, {and} \bibinfo{person}{Yea-Seul Kim}.} \bibinfo{year}{2023}\natexlab{}.
\newblock \showarticletitle{“Explain What a Treemap is”: Exploratory Investigation of Strategies for Explaining Unfamiliar Chart to Blind and Low Vision Users}. In \bibinfo{booktitle}{\emph{Proceedings of the ACM Conference on Human Factors in Computing Systems (CHI)}}. \bibinfo{publisher}{Association for Computing Machinery}, \bibinfo{address}{New York, NY, USA}, Article \bibinfo{articleno}{805}, \bibinfo{numpages}{13}~pages.
\newblock
\urldef\tempurl%
\url{https://doi.org/10.1145/3544548.3581139}
\showDOI{\tempurl}


\bibitem[Lall{\'e} et~al\mbox{.}(2016)]%
        {lalle16predicting}
\bibfield{author}{\bibinfo{person}{S{\'e}bastien Lall{\'e}}, \bibinfo{person}{Cristina Conati}, {and} \bibinfo{person}{Giuseppe Carenini}.} \bibinfo{year}{2016}\natexlab{}.
\newblock \showarticletitle{Predicting Confusion in Information Visualization from Eye Tracking and Interaction Data.}. In \bibinfo{booktitle}{\emph{Proceedings of the International Joint Conference on Artificial Intelligence (IJCAI)}}. \bibinfo{pages}{2529--2535}.
\newblock


\bibitem[Lam et~al\mbox{.}(2024)]%
        {lam2024lloom}
\bibfield{author}{\bibinfo{person}{Michelle~S. Lam}, \bibinfo{person}{Janice Teoh}, \bibinfo{person}{James~A. Landay}, \bibinfo{person}{Jeffrey Heer}, {and} \bibinfo{person}{Michael~S. Bernstein}.} \bibinfo{year}{2024}\natexlab{}.
\newblock \showarticletitle{Concept Induction: Analyzing Unstructured Text with High-Level Concepts Using LLooM}. In \bibinfo{booktitle}{\emph{Proceedings of the ACM Conference on Human Factors in Computing Systems (CHI)}}. \bibinfo{publisher}{Association for Computing Machinery}, \bibinfo{address}{New York, NY, USA}, Article \bibinfo{articleno}{766}, \bibinfo{numpages}{28}~pages.
\newblock
\urldef\tempurl%
\url{https://doi.org/10.1145/3613904.3642830}
\showDOI{\tempurl}


\bibitem[Latif et~al\mbox{.}(2022)]%
        {latif2022kori}
\bibfield{author}{\bibinfo{person}{Shahid Latif}, \bibinfo{person}{Zheng Zhou}, \bibinfo{person}{Yoon Kim}, \bibinfo{person}{Fabian Beck}, {and} \bibinfo{person}{Nam~Wook Kim}.} \bibinfo{year}{2022}\natexlab{}.
\newblock \showarticletitle{Kori: Interactive Synthesis of Text and Charts in Data Documents}.
\newblock \bibinfo{journal}{\emph{IEEE Transactions on Visualization and Computer Graphics}} \bibinfo{volume}{28}, \bibinfo{number}{1} (\bibinfo{date}{Jan} \bibinfo{year}{2022}), \bibinfo{pages}{184--194}.
\newblock
\urldef\tempurl%
\url{https://doi.org/10.1109/TVCG.2021.3114802}
\showDOI{\tempurl}


\bibitem[Lee(2024)]%
        {lee2024productivity}
\bibfield{author}{\bibinfo{person}{Geonsun Lee}.} \bibinfo{year}{2024}\natexlab{}.
\newblock \showarticletitle{Designing Interfaces for Enhanced Productivity and Collaboration in Virtual Workspaces}. In \bibinfo{booktitle}{\emph{Adjunct Proceedings of the IEEE International Symposium on Mixed and Augmented Reality}}. \bibinfo{pages}{666--667}.
\newblock
\urldef\tempurl%
\url{https://doi.org/10.1109/ISMAR-Adjunct64951.2024.00202}
\showDOI{\tempurl}


\bibitem[Lee et~al\mbox{.}(2025)]%
        {lee2025subg}
\bibfield{author}{\bibinfo{person}{Geonsun Lee}, \bibinfo{person}{Yue Yang}, \bibinfo{person}{Jennifer Healey}, {and} \bibinfo{person}{Dinesh Manocha}.} \bibinfo{year}{2025}\natexlab{}.
\newblock \showarticletitle{Since U Been Gone: Augmenting Context-Aware Transcriptions for Re-Engaging in Immersive VR Meetings}. In \bibinfo{booktitle}{\emph{Proceedings of the ACM Conference on Human Factors in Computing Systems}}. \bibinfo{publisher}{Association for Computing Machinery}, \bibinfo{address}{New York, NY, USA}, Article \bibinfo{articleno}{785}, \bibinfo{numpages}{20}~pages.
\newblock
\urldef\tempurl%
\url{https://doi.org/10.1145/3706598.3714078}
\showDOI{\tempurl}


\bibitem[Lee et~al\mbox{.}(2022)]%
        {lee2022coauthor}
\bibfield{author}{\bibinfo{person}{Mina Lee}, \bibinfo{person}{Percy Liang}, {and} \bibinfo{person}{Qian Yang}.} \bibinfo{year}{2022}\natexlab{}.
\newblock \showarticletitle{CoAuthor: Designing a Human-AI Collaborative Writing Dataset for Exploring Language Model Capabilities}. In \bibinfo{booktitle}{\emph{Proceedings of the ACM Conference on Human Factors in Computing Systems (CHI)}}. \bibinfo{publisher}{Association for Computing Machinery}, \bibinfo{address}{New York, NY, USA}, Article \bibinfo{articleno}{388}, \bibinfo{numpages}{19}~pages.
\newblock
\urldef\tempurl%
\url{https://doi.org/10.1145/3491102.3502030}
\showDOI{\tempurl}


\bibitem[Lee et~al\mbox{.}(2024)]%
        {lee2024paperweaver}
\bibfield{author}{\bibinfo{person}{Yoonjoo Lee}, \bibinfo{person}{Hyeonsu~B Kang}, \bibinfo{person}{Matt Latzke}, \bibinfo{person}{Juho Kim}, \bibinfo{person}{Jonathan Bragg}, \bibinfo{person}{Joseph~Chee Chang}, {and} \bibinfo{person}{Pao Siangliulue}.} \bibinfo{year}{2024}\natexlab{}.
\newblock \showarticletitle{PaperWeaver: Enriching Topical Paper Alerts by Contextualizing Recommended Papers with User-collected Papers}. In \bibinfo{booktitle}{\emph{Proceedings of the ACM Conference on Human Factors in Computing Systems (CHI)}}. \bibinfo{publisher}{Association for Computing Machinery}, \bibinfo{address}{New York, NY, USA}, Article \bibinfo{articleno}{19}, \bibinfo{numpages}{19}~pages.
\newblock
\urldef\tempurl%
\url{https://doi.org/10.1145/3613904.3642196}
\showDOI{\tempurl}


\bibitem[Li and Munby(1996)]%
        {li1996metacognitive}
\bibfield{author}{\bibinfo{person}{Shuyun Li} {and} \bibinfo{person}{Hugh Munby}.} \bibinfo{year}{1996}\natexlab{}.
\newblock \showarticletitle{Metacognitive strategies in second language academic reading: A qualitative investigation}.
\newblock \bibinfo{journal}{\emph{English for specific purposes}} \bibinfo{volume}{15}, \bibinfo{number}{3} (\bibinfo{year}{1996}), \bibinfo{pages}{199--216}.
\newblock


\bibitem[Li et~al\mbox{.}(2024)]%
        {li2024concernsinllmwriting}
\bibfield{author}{\bibinfo{person}{Zhuoyan Li}, \bibinfo{person}{Chen Liang}, \bibinfo{person}{Jing Peng}, {and} \bibinfo{person}{Ming Yin}.} \bibinfo{year}{2024}\natexlab{}.
\newblock \showarticletitle{The Value, Benefits, and Concerns of Generative AI-Powered Assistance in Writing}. In \bibinfo{booktitle}{\emph{Proceedings of the ACM Conference on Human Factors in Computing Systems (CHI)}}. \bibinfo{publisher}{Association for Computing Machinery}, \bibinfo{address}{New York, NY, USA}, Article \bibinfo{articleno}{1048}, \bibinfo{numpages}{25}~pages.
\newblock
\urldef\tempurl%
\url{https://doi.org/10.1145/3613904.3642625}
\showDOI{\tempurl}


\bibitem[Loshchilov and Hutter(2017)]%
        {loshchilov2017decoupled}
\bibfield{author}{\bibinfo{person}{Ilya Loshchilov} {and} \bibinfo{person}{Frank Hutter}.} \bibinfo{year}{2017}\natexlab{}.
\newblock \showarticletitle{Decoupled weight decay regularization}.
\newblock \bibinfo{journal}{\emph{arXiv preprint arXiv:1711.05101}} (\bibinfo{year}{2017}).
\newblock


\bibitem[Masson et~al\mbox{.}(2023)]%
        {masson2023statslator}
\bibfield{author}{\bibinfo{person}{Damien Masson}, \bibinfo{person}{Sylvain Malacria}, \bibinfo{person}{G\'{e}ry Casiez}, {and} \bibinfo{person}{Daniel Vogel}.} \bibinfo{year}{2023}\natexlab{}.
\newblock \showarticletitle{Statslator: Interactive Translation of NHST and Estimation Statistics Reporting Styles in Scientific Documents}. In \bibinfo{booktitle}{\emph{Proceedings of the ACM Symposium on User Interface Software and Technology (UIST)}}. \bibinfo{publisher}{Association for Computing Machinery}, \bibinfo{address}{New York, NY, USA}, Article \bibinfo{articleno}{91}, \bibinfo{numpages}{14}~pages.
\newblock
\urldef\tempurl%
\url{https://doi.org/10.1145/3586183.3606762}
\showDOI{\tempurl}


\bibitem[Naous et~al\mbox{.}(2024)]%
        {naous2024readme++}
\bibfield{author}{\bibinfo{person}{Tarek Naous}, \bibinfo{person}{Michael~J Ryan}, \bibinfo{person}{Anton Lavrouk}, \bibinfo{person}{Mohit Chandra}, {and} \bibinfo{person}{Wei Xu}.} \bibinfo{year}{2024}\natexlab{}.
\newblock \showarticletitle{Readme++: Benchmarking multilingual language models for multi-domain readability assessment}. In \bibinfo{booktitle}{\emph{Proceedings of the Conference on Empirical Methods in Natural Language Processing. Conference on Empirical Methods in Natural Language Processing}}, Vol.~\bibinfo{volume}{2024}. \bibinfo{pages}{12230}.
\newblock


\bibitem[of~Europe. Council for Cultural Co-operation. Education Committee. Modern Languages~Division(2001)]%
        {council2001common}
\bibfield{author}{\bibinfo{person}{Council of~Europe. Council for Cultural Co-operation. Education Committee. Modern Languages~Division}.} \bibinfo{year}{2001}\natexlab{}.
\newblock \bibinfo{booktitle}{\emph{Common European framework of reference for languages: Learning, teaching, assessment}}.
\newblock \bibinfo{publisher}{Cambridge University Press}.
\newblock


\bibitem[{Oxford University Press}({[n.\,d.]})]%
        {oxford5000}
\bibfield{author}{\bibinfo{person}{{Oxford University Press}}.} \bibinfo{year}{[n.\,d.]}\natexlab{}.
\newblock \bibinfo{title}{The Oxford 5000\texttrademark{} by CEFR Level}.
\newblock \bibinfo{howpublished}{\url{https://www.oxfordlearnersdictionaries.com/external/pdf/wordlists/oxford-3000-5000/The_Oxford_5000_by_CEFR_level.pdf}}.
\newblock
\newblock
\shownote{Oxford Learner's Dictionaries. Accessed: 2025-09-08}.


\bibitem[Palani et~al\mbox{.}(2023)]%
        {palani2023relatedly}
\bibfield{author}{\bibinfo{person}{Srishti Palani}, \bibinfo{person}{Aakanksha Naik}, \bibinfo{person}{Doug Downey}, \bibinfo{person}{Amy~X. Zhang}, \bibinfo{person}{Jonathan Bragg}, {and} \bibinfo{person}{Joseph~Chee Chang}.} \bibinfo{year}{2023}\natexlab{}.
\newblock \showarticletitle{Relatedly: Scaffolding Literature Reviews with Existing Related Work Sections}. In \bibinfo{booktitle}{\emph{Proceedings of the ACM Conference on Human Factors in Computing Systems (CHI)}}. \bibinfo{publisher}{Association for Computing Machinery}, \bibinfo{address}{New York, NY, USA}, Article \bibinfo{articleno}{742}, \bibinfo{numpages}{20}~pages.
\newblock
\urldef\tempurl%
\url{https://doi.org/10.1145/3544548.3580841}
\showDOI{\tempurl}


\bibitem[Parsons and Shukla(2020)]%
        {parsons20chartjunk}
\bibfield{author}{\bibinfo{person}{Paul Parsons} {and} \bibinfo{person}{Prakash Shukla}.} \bibinfo{year}{2020}\natexlab{}.
\newblock \showarticletitle{Data Visualization Practitioners’ Perspectives on Chartjunk}. In \bibinfo{booktitle}{\emph{Short Proceedings of the IEEE Visualization Conference}}. \bibinfo{pages}{211--215}.
\newblock
\urldef\tempurl%
\url{https://doi.org/10.1109/VIS47514.2020.00049}
\showDOI{\tempurl}


\bibitem[Peng et~al\mbox{.}(2022)]%
        {peng2022crebot}
\bibfield{author}{\bibinfo{person}{Zhenhui Peng}, \bibinfo{person}{Yuzhi Liu}, \bibinfo{person}{Hanqi Zhou}, \bibinfo{person}{Zuyu Xu}, {and} \bibinfo{person}{Xiaojuan Ma}.} \bibinfo{year}{2022}\natexlab{}.
\newblock \showarticletitle{CReBot: Exploring interactive question prompts for critical paper reading}.
\newblock \bibinfo{journal}{\emph{International Journal of Human-Computer Studies}}  \bibinfo{volume}{167} (\bibinfo{year}{2022}), \bibinfo{pages}{102898}.
\newblock
\urldef\tempurl%
\url{https://doi.org/10.1016/j.ijhcs.2022.102898}
\showDOI{\tempurl}


\bibitem[Ramadhianti and Somba(2023)]%
        {ramadhianti2023reading}
\bibfield{author}{\bibinfo{person}{Agustina Ramadhianti} {and} \bibinfo{person}{Sugianti Somba}.} \bibinfo{year}{2023}\natexlab{}.
\newblock \showarticletitle{Reading comprehension difficulties in Indonesian EFL students}.
\newblock \bibinfo{journal}{\emph{Journal of English Language Teaching and Literature (JELTL)}} \bibinfo{volume}{6}, \bibinfo{number}{1} (\bibinfo{year}{2023}), \bibinfo{pages}{1--11}.
\newblock


\bibitem[Ramanathan and Atkinson(1999)]%
        {ramanathan1999individualism}
\bibfield{author}{\bibinfo{person}{Vai Ramanathan} {and} \bibinfo{person}{Dwight Atkinson}.} \bibinfo{year}{1999}\natexlab{}.
\newblock \showarticletitle{Individualism, academic writing, and ESL writers}.
\newblock \bibinfo{journal}{\emph{Journal of second language writing}} \bibinfo{volume}{8}, \bibinfo{number}{1} (\bibinfo{year}{1999}), \bibinfo{pages}{45--75}.
\newblock


\bibitem[Shin et~al\mbox{.}(2024)]%
        {shin24situatedsuvey}
\bibfield{author}{\bibinfo{person}{Sungbok Shin}, \bibinfo{person}{Andrea Batch}, \bibinfo{person}{Peter William~Scott Butcher}, \bibinfo{person}{Panagiotis~D. Ritsos}, {and} \bibinfo{person}{Niklas Elmqvist}.} \bibinfo{year}{2024}\natexlab{}.
\newblock \showarticletitle{The Reality of the Situation: A Survey of Situated Analytics}.
\newblock \bibinfo{journal}{\emph{IEEE Transactions on Visualization and Computer Graphics}} \bibinfo{volume}{30}, \bibinfo{number}{8} (\bibinfo{year}{2024}), \bibinfo{pages}{5147--5164}.
\newblock
\urldef\tempurl%
\url{https://doi.org/10.1109/TVCG.2023.3285546}
\showDOI{\tempurl}


\bibitem[Shin et~al\mbox{.}(2025a)]%
        {shin2025visualizationary}
\bibfield{author}{\bibinfo{person}{Sungbok Shin}, \bibinfo{person}{Sanghyun Hong}, {and} \bibinfo{person}{Niklas Elmqvist}.} \bibinfo{year}{2025}\natexlab{a}.
\newblock \showarticletitle{Visualizationary: Automating Design Feedback for Visualization Designers Using LLMs}.
\newblock \bibinfo{journal}{\emph{IEEE Transactions on Visualization and Computer Graphics}} (\bibinfo{year}{2025}), \bibinfo{pages}{1--17}.
\newblock
\urldef\tempurl%
\url{https://doi.org/10.1109/TVCG.2025.3579700}
\showDOI{\tempurl}


\bibitem[Shin et~al\mbox{.}(2025b)]%
        {shin25drillboards}
\bibfield{author}{\bibinfo{person}{Sungbok Shin}, \bibinfo{person}{Inyoup Na}, {and} \bibinfo{person}{Niklas Elmqvist}.} \bibinfo{year}{2025}\natexlab{b}.
\newblock \showarticletitle{Drillboards: Adaptive Visualization Dashboards for Dynamic Personalization of Visualization Experiences}.
\newblock \bibinfo{journal}{\emph{IEEE Transactions on Visualization and Computer Graphics}} (\bibinfo{year}{2025}), \bibinfo{pages}{1--14}.
\newblock
\urldef\tempurl%
\url{https://doi.org/10.1109/TVCG.2025.3542606}
\showDOI{\tempurl}


\bibitem[Sims and Conati(2020)]%
        {sims20userconfusion}
\bibfield{author}{\bibinfo{person}{Shane~D. Sims} {and} \bibinfo{person}{Cristina Conati}.} \bibinfo{year}{2020}\natexlab{}.
\newblock \showarticletitle{A Neural Architecture for Detecting User Confusion in Eye-tracking Data}. In \bibinfo{booktitle}{\emph{Proceedings of the International Conference on Multimodal Interaction}}. \bibinfo{publisher}{Association for Computing Machinery}, \bibinfo{address}{New York, NY, USA}, \bibinfo{pages}{15–23}.
\newblock
\urldef\tempurl%
\url{https://doi.org/10.1145/3382507.3418828}
\showDOI{\tempurl}


\bibitem[Singh et~al\mbox{.}(2024)]%
        {singh2024figura11y}
\bibfield{author}{\bibinfo{person}{Nikhil Singh}, \bibinfo{person}{Lucy~Lu Wang}, {and} \bibinfo{person}{Jonathan Bragg}.} \bibinfo{year}{2024}\natexlab{}.
\newblock \showarticletitle{FigurA11y: AI Assistance for Writing Scientific Alt Text}. In \bibinfo{booktitle}{\emph{Proceedings of the International Conference on Intelligent User Interfaces (IUI)}}. \bibinfo{publisher}{Association for Computing Machinery}, \bibinfo{address}{New York, NY, USA}, \bibinfo{pages}{886–906}.
\newblock
\urldef\tempurl%
\url{https://doi.org/10.1145/3640543.3645212}
\showDOI{\tempurl}


\bibitem[Stokes et~al\mbox{.}(2023)]%
        {stokes2023balancetextchart}
\bibfield{author}{\bibinfo{person}{Chase Stokes}, \bibinfo{person}{Vidya Setlur}, \bibinfo{person}{Bridget Cogley}, \bibinfo{person}{Arvind Satyanarayan}, {and} \bibinfo{person}{Marti~A. Hearst}.} \bibinfo{year}{2023}\natexlab{}.
\newblock \showarticletitle{Striking a Balance: Reader Takeaways and Preferences when Integrating Text and Charts}.
\newblock \bibinfo{journal}{\emph{IEEE Transactions on Visualization and Computer Graphics}} \bibinfo{volume}{29}, \bibinfo{number}{1} (\bibinfo{date}{Jan} \bibinfo{year}{2023}), \bibinfo{pages}{1233--1243}.
\newblock
\urldef\tempurl%
\url{https://doi.org/10.1109/TVCG.2022.3209383}
\showDOI{\tempurl}


\bibitem[Strobelt et~al\mbox{.}(2016)]%
        {strobelt16highlighttext}
\bibfield{author}{\bibinfo{person}{Hendrik Strobelt}, \bibinfo{person}{Daniela Oelke}, \bibinfo{person}{Bum~Chul Kwon}, \bibinfo{person}{Tobias Schreck}, {and} \bibinfo{person}{Hanspeter Pfister}.} \bibinfo{year}{2016}\natexlab{}.
\newblock \showarticletitle{Guidelines for Effective Usage of Text Highlighting Techniques}.
\newblock \bibinfo{journal}{\emph{IEEE Transactions on Visualization and Computer Graphics}} \bibinfo{volume}{22}, \bibinfo{number}{1} (\bibinfo{year}{2016}), \bibinfo{pages}{489--498}.
\newblock
\urldef\tempurl%
\url{https://doi.org/10.1109/TVCG.2015.2467759}
\showDOI{\tempurl}


\bibitem[Tse and Sahasrabudhe({[n.\,d.]})]%
        {o_levels}
\bibfield{author}{\bibinfo{person}{Emily Tse} {and} \bibinfo{person}{Ujjaini Sahasrabudhe}.} \bibinfo{year}{[n.\,d.]}\natexlab{}.
\newblock \bibinfo{title}{O's and A's and other Letters of the Alphabet: Making Sense of British-based Secondary and Pre-University Level Qualitifications}.
\newblock
\newblock
\newblock
\shownote{accessed on: April 2025}.


\bibitem[Uchida and Negishi(2018)]%
        {uchida2018assigning}
\bibfield{author}{\bibinfo{person}{Satoru Uchida} {and} \bibinfo{person}{Masashi Negishi}.} \bibinfo{year}{2018}\natexlab{}.
\newblock \showarticletitle{Assigning CEFR-J levels to English texts based on textual features}. In \bibinfo{booktitle}{\emph{Proceedings of Asia Pacific Corpus Linguistics Conference}}, Vol.~\bibinfo{volume}{4}. \bibinfo{pages}{463--467}.
\newblock


\bibitem[Wambsganss et~al\mbox{.}(2021)]%
        {wamgsganss2021arguetutor}
\bibfield{author}{\bibinfo{person}{Thiemo Wambsganss}, \bibinfo{person}{Tobias Kueng}, \bibinfo{person}{Matthias Soellner}, {and} \bibinfo{person}{Jan~Marco Leimeister}.} \bibinfo{year}{2021}\natexlab{}.
\newblock \showarticletitle{ArgueTutor: An Adaptive Dialog-Based Learning System for Argumentation Skills}. In \bibinfo{booktitle}{\emph{Proceedings of the ACM Conference on Human Factors in Computing Systems (CHI)}}. \bibinfo{publisher}{Association for Computing Machinery}, \bibinfo{address}{New York, NY, USA}, Article \bibinfo{articleno}{683}, \bibinfo{numpages}{13}~pages.
\newblock
\urldef\tempurl%
\url{https://doi.org/10.1145/3411764.3445781}
\showDOI{\tempurl}


\bibitem[Wang et~al\mbox{.}(2024)]%
        {wang2024chatprcs}
\bibfield{author}{\bibinfo{person}{Xizhe Wang}, \bibinfo{person}{Yihua Zhong}, \bibinfo{person}{Changqin Huang}, {and} \bibinfo{person}{Xiaodi Huang}.} \bibinfo{year}{2024}\natexlab{}.
\newblock \showarticletitle{ChatPRCS: A Personalized Support System for English Reading Comprehension Based on ChatGPT}.
\newblock \bibinfo{journal}{\emph{IEEE Transactions on Learning Technologies}}  \bibinfo{volume}{17} (\bibinfo{year}{2024}), \bibinfo{pages}{1762--1776}.
\newblock
\urldef\tempurl%
\url{https://doi.org/10.1109/TLT.2024.3405747}
\showDOI{\tempurl}


\bibitem[Yang et~al\mbox{.}(2025)]%
        {qwen3}
\bibfield{author}{\bibinfo{person}{An Yang}, \bibinfo{person}{Anfeng Li}, \bibinfo{person}{Baosong Yang}, \bibinfo{person}{Beichen Zhang}, \bibinfo{person}{Binyuan Hui}, \bibinfo{person}{Bo Zheng}, \bibinfo{person}{Bowen Yu}, \bibinfo{person}{Chang Gao}, \bibinfo{person}{Chengen Huang}, \bibinfo{person}{Chenxu Lv}, \bibinfo{person}{Chujie Zheng}, \bibinfo{person}{Dayiheng Liu}, \bibinfo{person}{Fan Zhou}, \bibinfo{person}{Fei Huang}, \bibinfo{person}{Feng Hu}, \bibinfo{person}{Hao Ge}, \bibinfo{person}{Haoran Wei}, \bibinfo{person}{Huan Lin}, \bibinfo{person}{Jialong Tang}, \bibinfo{person}{Jian Yang}, \bibinfo{person}{Jianhong Tu}, \bibinfo{person}{Jianwei Zhang}, \bibinfo{person}{Jianxin Yang}, \bibinfo{person}{Jiaxi Yang}, \bibinfo{person}{Jing Zhou}, \bibinfo{person}{Jingren Zhou}, \bibinfo{person}{Junyang Lin}, \bibinfo{person}{Kai Dang}, \bibinfo{person}{Keqin Bao}, \bibinfo{person}{Kexin Yang}, \bibinfo{person}{Le Yu}, \bibinfo{person}{Lianghao Deng}, \bibinfo{person}{Mei Li}, \bibinfo{person}{Mingfeng
  Xue}, \bibinfo{person}{Mingze Li}, \bibinfo{person}{Pei Zhang}, \bibinfo{person}{Peng Wang}, \bibinfo{person}{Qin Zhu}, \bibinfo{person}{Rui Men}, \bibinfo{person}{Ruize Gao}, \bibinfo{person}{Shixuan Liu}, \bibinfo{person}{Shuang Luo}, \bibinfo{person}{Tianhao Li}, \bibinfo{person}{Tianyi Tang}, \bibinfo{person}{Wenbiao Yin}, \bibinfo{person}{Xingzhang Ren}, \bibinfo{person}{Xinyu Wang}, \bibinfo{person}{Xinyu Zhang}, \bibinfo{person}{Xuancheng Ren}, \bibinfo{person}{Yang Fan}, \bibinfo{person}{Yang Su}, \bibinfo{person}{Yichang Zhang}, \bibinfo{person}{Yinger Zhang}, \bibinfo{person}{Yu Wan}, \bibinfo{person}{Yuqiong Liu}, \bibinfo{person}{Zekun Wang}, \bibinfo{person}{Zeyu Cui}, \bibinfo{person}{Zhenru Zhang}, \bibinfo{person}{Zhipeng Zhou}, {and} \bibinfo{person}{Zihan Qiu}.} \bibinfo{year}{2025}\natexlab{}.
\newblock \showarticletitle{Qwen3 Technical Report}.
\newblock \bibinfo{journal}{\emph{arXiv preprint arXiv:2505.09388}} (\bibinfo{year}{2025}).
\newblock


\bibitem[Yuan et~al\mbox{.}(2023)]%
        {yuan2023critrainer}
\bibfield{author}{\bibinfo{person}{Kangyu Yuan}, \bibinfo{person}{Hehai Lin}, \bibinfo{person}{Shilei Cao}, \bibinfo{person}{Zhenhui Peng}, \bibinfo{person}{Qingyu Guo}, {and} \bibinfo{person}{Xiaojuan Ma}.} \bibinfo{year}{2023}\natexlab{}.
\newblock \showarticletitle{CriTrainer: An Adaptive Training Tool for Critical Paper Reading}. In \bibinfo{booktitle}{\emph{Proceedings of the ACM Symposium on User Interface Software and Technology (UIST)}}. \bibinfo{publisher}{Association for Computing Machinery}, \bibinfo{address}{New York, NY, USA}, Article \bibinfo{articleno}{44}, \bibinfo{numpages}{17}~pages.
\newblock
\urldef\tempurl%
\url{https://doi.org/10.1145/3586183.3606816}
\showDOI{\tempurl}


\bibitem[Zamel(1982)]%
        {zamel1982writing}
\bibfield{author}{\bibinfo{person}{Vivian Zamel}.} \bibinfo{year}{1982}\natexlab{}.
\newblock \showarticletitle{Writing: The process of discovering meaning}.
\newblock \bibinfo{journal}{\emph{TESOL quarterly}} \bibinfo{volume}{16}, \bibinfo{number}{2} (\bibinfo{year}{1982}), \bibinfo{pages}{195--209}.
\newblock


\bibitem[Zhang et~al\mbox{.}(2023)]%
        {zhang2023concepteva}
\bibfield{author}{\bibinfo{person}{Xiaoyu Zhang}, \bibinfo{person}{Jianping Li}, \bibinfo{person}{Po-Wei Chi}, \bibinfo{person}{Senthil Chandrasegaran}, {and} \bibinfo{person}{Kwan-Liu Ma}.} \bibinfo{year}{2023}\natexlab{}.
\newblock \showarticletitle{ConceptEVA: Concept-Based Interactive Exploration and Customization of Document Summaries}. In \bibinfo{booktitle}{\emph{Proceedings of the ACM Conference on Human Factors in Computing Systems (CHI)}}. \bibinfo{publisher}{Association for Computing Machinery}, \bibinfo{address}{New York, NY, USA}, Article \bibinfo{articleno}{204}, \bibinfo{numpages}{16}~pages.
\newblock
\showISBNx{9781450394215}
\urldef\tempurl%
\url{https://doi.org/10.1145/3544548.3581260}
\showDOI{\tempurl}


\bibitem[Zyska et~al\mbox{.}(2023)]%
        {zyska2023care}
\bibfield{author}{\bibinfo{person}{Dennis Zyska}, \bibinfo{person}{Nils Dycke}, \bibinfo{person}{Jan Buchmann}, \bibinfo{person}{Ilia Kuznetsov}, {and} \bibinfo{person}{Iryna Gurevych}.} \bibinfo{year}{2023}\natexlab{}.
\newblock \showarticletitle{{CARE}: Collaborative {AI}-Assisted Reading Environment}. In \bibinfo{booktitle}{\emph{Proceedings of the Annual Meeting of the Association for Computational Linguistics (Volume 3: System Demonstrations)}}. \bibinfo{publisher}{Association for Computational Linguistics}, \bibinfo{address}{Toronto, Canada}, \bibinfo{pages}{291--303}.
\newblock
\urldef\tempurl%
\url{https://doi.org/10.18653/v1/2023.acl-demo.28}
\showDOI{\tempurl}


\end{thebibliography}

\end{document}